\documentclass{lmcs}

\keywords{%
lambda calculus,
program approximation,
linear approximation,
infinitary rewriting,
algebraic rewriting,
conservativity,
quantitative semantics%
}

\usepackage[T1]{fontenc}

\usepackage{alphabeta}
\DeclareUnicodeCharacter{2264}{\leq}
\DeclareUnicodeCharacter{2265}{\geq}

\usepackage{array}
\usepackage{adjustbox}
\usepackage{amssymb}
\usepackage{booktabs}
\usepackage{csquotes}
\usepackage{float}
\usepackage{needspace}
\usepackage{stackengine}
\usepackage{todonotes}
\usepackage{xcolor}
\usepackage{xspace}

\usepackage[inline]{enumitem}
\NewDocumentEnvironment{ienumerate}{}
	{\begin{enumerate*}[label={(\roman*)}]}
	{\end{enumerate*}}

\usepackage{ebproof}
\NewDocumentCommand \proofrule {m} {\ensuremath{\smash{\mathrm{(#1)}}}}
\ebproofset{
	center=false,
	right label template=\small\proofrule{\inserttext},
}

\usepackage[style=alphabetic]{biblatex}
\bibliography{conservativity.bib}

\theoremstyle{plain}
\newtheorem{simulthm}[thm]{Simulation theorem}
\newtheorem{simulcor}[thm]{Simulation corollary}
\newtheorem{conservthm}[thm]{Conservativity theorem}
\newtheorem{conservcor}[thm]{Conservativity corollary}

\AtEndPreamble{
\usepackage[capitalise]{cleveref}
\crefname{cor}{Corollary}{Corollaries}
\crefname{defi}{Definition}{Definitions}
\crefname{lem}{Lemma}{Lemmas}
\crefname{prob}{Problem}{Problems}
\crefname{thm}{Theorem}{Theorems}
\crefname{nota}{Notation}{Notations}
\crefname{obs}{Observation}{Observations}
\crefname{conj}{Conjecture}{Conjectures}

\crefname{simulthm}{Theorem}{Theorems}
\crefname{simulcor}{Corollary}{Corollaries}
\crefname{conservthm}{Theorem}{Theorems}
\crefname{conservcor}{Corollary}{Corollaries}
}

\colorlet{firedredex}{black!40!blue!10}
\NewDocumentCommand \defemph 	{m}		{\textbf{#1}}
\NewDocumentCommand \ie 		{} 		{\emph{i.e.}\@\xspace}
\NewDocumentCommand \eg 		{} 		{\emph{e.g.}\@\xspace}
\NewDocumentCommand \wrt 		{} 		{wrt.\@\xspace}

\AtEndPreamble{
\RenewDocumentCommand \SS 		{}		{ \mathbf{S} }
}
\NewDocumentCommand \TT 		{}		{ \mathbf{T} }
\NewDocumentCommand \UU 		{}		{ \mathbf{U} }
\NewDocumentCommand \VV 		{}		{ \mathbf{V} }

\NewDocumentCommand \Nats 		{}		{ \mathbb{N} }
\NewDocumentCommand \Rationals 	{}		{ \mathbb{Q} }
\NewDocumentCommand \SSS 		{}		{ \mathbb{S} }

\NewDocumentCommand \eqdef 		{}		{ \coloneqq }

\newlength{\prooftreevsep}
\NewDocumentEnvironment{prooftreeset}{s}%
	{	\vspace{\prooftreevsep}%
		\IfBooleanTF{#1}{\gather}{\csname gather*\endcsname} }%
	{	\IfBooleanTF{#1}{\endgather}{\csname endgather*\endcsname}%
		\vspace{\prooftreevsep}}

\NewDocumentEnvironment {widecenter} {}
	{\begin{adjustbox}{center,max width=\paperwidth}}
	{\end{adjustbox}}

\NewDocumentCommand \tunderline		{m}
    {\tikz[baseline=(foo.base)]{
        \node[inner xsep=0pt,inner ysep=2pt,outer sep=0pt] (foo) {#1};
        \draw (foo.south west) -- (foo.south east);
    }}
\ebproofnewstyle{doubleprem}{
	template={\tunderline{$\strut\inserttext$}}
}
\NewDocumentCommand \tunderdash		{m}
    {\tikz[baseline=(foo.base)]{
        \node[inner xsep=0pt,inner ysep=2pt,outer sep=0pt] (foo) {#1};
        \draw[densely dashed] (foo.south west) -- (foo.south east);
    }}

\NewDocumentCommand \set 		{mo}
	{ \left\{ #1 \IfValueT{#2}{ \,\middle\vert\, #2 } \right\} }
\NewDocumentCommand \ms 		{m}		{ \bar{#1} }

\NewDocumentCommand \perm		{}		{ \mathfrak{S} }
\NewDocumentCommand \red 		{oo}	{
	\longrightarrow\IfValueT{#1}{_{#1}}\IfValueT{#2}{^{#2}}
	}
\NewDocumentCommand	 \reds		{O{}}	{ \red[#1][*] }
\NewDocumentCommand	 \redi		{O{}}	{ \red[#1][001] }

\NewDocumentCommand \letrelsbreathe {}
	{\thickmuskip=12mu plus 5mu}
\newlength{\widerelwidth}
\NewDocumentCommand \widerel 	{mm}	{
	\settowidth{\widerelwidth}{\ensuremath{#1}}
	\mathrel{\makebox[\widerelwidth][c]{\ensuremath{#2}}}
	}

\NewDocumentCommand \Vars 		{}		{ \mathcal{V} }
\NewDocumentCommand \lc 		{}		{ \Lambda }
\NewDocumentCommand \lci 		{}		{ \Lambda^{001} }
\NewDocumentCommand \lbc 		{}		{ \Lambda_{\bot} }
\NewDocumentCommand \lbci 		{}		{ \Lambda_{\bot}^{001} }

\NewDocumentCommand \subst 		{mmo}	{ #1[#2/\IfValueTF{#3}{#3}{x}] }

\NewDocumentCommand \bred 		{o}		{ \red[β\IfValueT{#1}{#1}] }
\NewDocumentCommand \breds 		{o}		{ \red[β\IfValueT{#1}{#1}][*] }
\NewDocumentCommand \bredi 		{o}		{ \red[β\IfValueT{#1}{#1}][001] }

\NewDocumentCommand \botred 	{o}		{ \red[\bot\IfValueT{#1}{#1}] }
\NewDocumentCommand \botreds 	{o}		{ \red[\bot\IfValueT{#1}{#1}][*] }
\NewDocumentCommand \botredi 	{o}		{ \red[\bot\IfValueT{#1}{#1}][001] }

\NewDocumentCommand \bbotred 	{o}		{ \red[β\bot\IfValueT{#1}{#1}] }
\NewDocumentCommand \bbotreds 	{o}		{ \red[β\bot\IfValueT{#1}{#1}][*] }
\NewDocumentCommand \bbotredi 	{o}		{ \red[β\bot\IfValueT{#1}{#1}][001] }

\NewDocumentCommand \yfp 		{o}		{ \mathtt{Y}\IfValueT{#1}{_{\!#1}} }
\NewDocumentCommand \infrapp	{m}		{ (#1)^{\omega} }
\NewDocumentCommand \deltadelta {}		{ \text{\texttt{Ω}} }
\DeclareMathOperator \BT 				{ BT }

\NewDocumentCommand \resource 	{}		{ \mathrm{r} }

\NewDocumentCommand \rapp 		{smm}	{ 
	\left( #2 \right) \IfBooleanTF{#1}{ #3 }{ \ms{#3} }
	}

\NewDocumentCommand \rc 		{ss}	{ \mathord{
	\IfBooleanT{#2}{(}\IfBooleanT{#1}{!}\IfBooleanT{#2}{)}
	\Lambda_{\resource}
	}}

\NewDocumentCommand \frsums 	{sso}	{ 
	\IfValueTF{#3}{#3}{\Nats}^%
	{ (\IfBooleanTF{#2}{\rc**}{\IfBooleanTF{#1}{\rc*}{\rc}}) } }
\NewDocumentCommand \irsums 	{sso}	{ 
	\IfValueTF{#3}{#3}{\SSS}^%
	{ \IfBooleanTF{#2}{\rc**}{\IfBooleanTF{#1}{\rc*}{\rc}} } }
\NewDocumentCommand \sumsupp 	{m} 	{ \left|#1\right| }
\NewDocumentCommand \suminclusion {}	{ \subseteq }

\NewDocumentCommand \rsubst 	{smmo}	{%
	#2\langle \IfBooleanTF{#1}{#3}{\ms{#3}} / \IfValueTF{#4}{#4}{x} \rangle
	}
\NewDocumentCommand \bigrsubst 	{mmo}	{%
	#1\left\langle #2 \middle/ \IfValueTF{#3}{#3}{x} \right\rangle
	}

\DeclareMathOperator \rdepth			{depth}
\NewDocumentCommand \abovedepth {som}	{
	\IfBooleanT{#1}{\left}\lfloor #3 \IfBooleanT{#1}{\right}\rfloor
	_{\IfValueTF{#2}{#2}{d}} }

\NewDocumentCommand \longrightharpoonup {}{ \relbar\joinrel\rightharpoonup }
\NewDocumentCommand \Orred 		{o}		{
	\longrightharpoonup_{\resource\IfValueT{#1}{#1}} }
\NewDocumentCommand \Orredr 	{o}		{ 
	\Orred[\IfValueT{#1}{#1}]^? }

\NewDocumentCommand \rred 		{o}		{
	\red[\resource\IfValueT{#1}{#1}] }
\NewDocumentCommand \rredr 		{o}		{ 
	\red[\resource\IfValueT{#1}{#1}][?] }
\NewDocumentCommand \rreds 		{o}		{ 
	\red[\resource\IfValueT{#1}{#1}][*] }

\NewDocumentCommand \longtwoheadrightarrow {}{ 
	\relbar\joinrel\twoheadrightarrow }
\NewDocumentCommand \Lrred 		{o}		{
	\longtwoheadrightarrow_{\resource\IfValueT{#1}{#1}} }
\DeclareMathOperator \nf 				{ nf }

\NewDocumentCommand \Tay 		{}		{ \mathcal{T} }

\NewDocumentCommand \mashup 	{o}		{
	\IfValueTF{#1}{\vdash^{#1}}{\vdash} }

\NewDocumentCommand \lctorc 	{mo}	{
	\lfloor #1 \rfloor_{\resource\IfValueT{#2}{,#2}} }

\NewDocumentCommand \idterm 	{}		{ \mathtt{I} }
\NewDocumentCommand \ttterm 	{}		{ \mathtt{T} }
\NewDocumentCommand \ffterm 	{}		{ \mathtt{F} }
\NewDocumentCommand \churchn 	{o}		{ \mathtt{\IfValueTF{#1}{#1}{n}} }
\NewDocumentCommand \churchsucc {}		{ \mathtt{Succ} }
\NewDocumentCommand \appterm 	{m}		{ \langle #1 \rangle }

\NewDocumentCommand \lamP 		{}		{ \mathtt{P} }
\NewDocumentCommand \lamQ 		{}		{ \mathtt{Q} }
\NewDocumentCommand \acc 		{s}		{
	\IfBooleanTF{#1}{ \bar{\mathtt{A}} }{ \mathtt{A} } }

\newcommand*{\fired}[1]{%
  \tikz[baseline=(X.base)] \node[rectangle, fill=firedredex, 
  rounded corners, inner sep=0.5mm] (X) {\ensuremath{#1}};%
}

\NewDocumentCommand \hred 		{}		{ \red[\mathrm h] }
\NewDocumentCommand \hreds 		{}		{ \red[\mathrm h][*] }
\NewDocumentCommand \ired 		{}		{ \red[\mathrm i] }
\NewDocumentCommand \ireds 		{}		{ \red[\mathrm i][*] }

\NewDocumentCommand \coh 		{}		{
	\mathrel{\ensurestackMath{\stackanchor[0ex]{\frown}{\smile}}} }

\NewDocumentCommand \cohresource {s}
	{ \ensurestackMath{\stackon
	{\smash{\IfBooleanF{#1}{\scriptstyle}\resource}%
		\IfBooleanT{#1}{\vphantom{\scalebox{.5}{x}}}}
	{\IfBooleanTF{#1}{\scriptstyle}{\scriptscriptstyle}\frown}} }

\NewDocumentCommand \cohlift 	{mo}	{ \mathrel{\ensurestackMath{
	\stackon[-.2ex]{#1}{\smash{ \frown\IfValueT{#2}{\kern#2} }} }}}

\NewDocumentCommand \cohOrred 	{o}		{
	\cohlift{\longrightharpoonup}[.2em]_{\resource\IfValueT{#1}{#1}} }
\NewDocumentCommand \cohrred 	{o}		{ 
	\cohlift{\red}[.1em]_{\resource\IfValueT{#1}{#1}} }
\NewDocumentCommand \cohrreds 	{o}		{ 
	\cohlift{\red}[.1em]_{\resource\IfValueT{#1}{#1}}^* }
\NewDocumentCommand \cohrredi 	{o}		{ 
	\cohlift{\red}[.1em]_{\resource\IfValueT{#1}{#1}}^{\infty} }

\NewDocumentCommand \cohOrbotred 	{o}		{
	\cohOrred[\bot\IfValueT{#1}{#1}] }
\NewDocumentCommand \cohrbotred 	{o}		{ 
	\cohrred[\bot\IfValueT{#1}{#1}] }
\NewDocumentCommand \cohrbotreds 	{o}		{ 
	\cohrreds[\bot\IfValueT{#1}{#1}] }
\NewDocumentCommand \cohrbotredi 	{o}		{ 
	\cohrredi[\bot\IfValueT{#1}{#1}] }

\usepackage{hyperref}

\begin{document}

% If the title is longer than 55 characters, then specify a shorter running title as the optional argument to \title. The running title should be roughyl at most 55 characters:
\title[How to play the Accordion]
	{How to play the Accordion\texorpdfstring{:\\}{. }%
	Uniformity and the (non-)conservativity of the linear approximation
	of the \texorpdfstring{\MakeLowercase{λ}}{λ}-calculus\rsuper*}
\titlecomment{{\lsuper*} Improved and extended version of the article 
	\cite{Cerda.Vau.25}
	published in the proceedings of the 42nd~International Symposium on 
	Theoretical Aspects of Computer Science (STACS 2025).}
%\thanks{}

% affiliations are numbered automatically with a, b, c (see below)
% use the optional argument to indicate the affiliation(s) of each author
% omit the argument if there is only one author, or only one affiliation
\author[R.~Cerda]{Rémy Cerda\lmcsorcid{0000-0003-0731-6211}}[a,b,c]
\thanks{The first author was partially funded by the French ANR project 
	RECIPROG (ANR-21-CE48-019).}

\author[L.~Vaux Auclair]
	{Lionel Vaux Auclair\lmcsorcid{0000-0001-9466-418X}}[a]
\thanks{The second author was partially supported by the French ANR projects 
	LambdaComb (ANR-21-CE48-0017), RECIPROG (ANR-21-CE48-019),
	and PPS (ANR-19-CE48-0014).}

\address{Aix-Marseille Université, CNRS, I2M, France}
\email{Remy.Cerda@math.cnrs.fr, Lionel.Vaux@math.cnrs.fr}
\address{Université Paris Cité, CNRS, IRIF, F-75013, Paris, France}
\address{Università di Bologna, Italy}

%% --------------------------------------------------------------------------

\begin{abstract}
	Twenty years after its introduction by Ehrhard and Regnier,
	differentiation in λ-calculus and in linear logic is now
	a celebrated tool.
	In particular, it allows to establish a Taylor expansion formula
	for various λ-calculi, hence providing a theory of
	linear approximations for these calculi.
	In the pure λ-calculus, 
        the linear approximants of λ-terms supporting this Taylor expansion
        are the terms of a so-called resource calculus, which is
        equipped with a finitary (strongly normalising) reduction;
        and the efficiency of this linear approximation is expressed
	by results stating that the (possibly) infinitary β-reduction of λ-terms
	is simulated by the reduction of their Taylor expansions,
        which is induced by the iterated reduction of resource terms.
	In terms of rewriting systems, resource reduction
	(operating on infinite linear combinations of Taylor approximants)
	is an extension of β-reduction.
	
	In this article, we address the converse property, conservativity:
	do all reductions between Taylor expansions
	arise from actual β-reductions?
	We show that if we restrict the setting to finite terms and β-reduction
	sequences, then the linear approximation is conservative.
	However, as soon as one allows infinitary reduction sequences
	this property is broken. We design a counter-example, the Accordion.
	Then we show how restricting the reduction of the Taylor approximants
	allows to build a conservative extension of the β-reduction
	preserving good simulation properties;
	this restriction relies on uniformity,
	a property that was already at the core
	of Ehrhard and Regnier's pioneering work.
	Finally, we extend our work to β$\bot$-reductions,
	which play a key role in λ-calculus as they relate a λ-term
	to its Böhm tree.
\end{abstract}

\maketitle

\needspace{3cm}
\tableofcontents

%% --------------------------------------------------------------------------

\section{Introduction}

The traditional approach to program approximation
in a functional setting consists in describing the
total information that a (potentially non-terminating) program can produce 
as the supremum of the finite pieces of information
it can produce in finite time.
This idea of a \emph{continuous} approximation is at the core of
the Scott semantics of λ-calculi \cite{Scott93},
and can be formulated in syntactic terms by showing that
the Böhm tree of a λ-term
is the limit of the approximants produced by hereditary head reduction
\cite{Hyland76,Wadsworth78,Barendregt84}.

More recently, Ehrhard and Regnier introduced the differential λ-calculus
and differential linear logic
\cite{EhrhardRegnier03,EhrhardRegnier05},
following ideas rooted in the semantics of linear logic
\cite{Girard87,Ehrhard02,Ehrhard05}.
This suggested the renewed approach of \emph{linear} approximation
of functional programs.
In this setting,
a program (\ie a λ-term) is approximated by multilinear 
(or \enquote{polynomial}) programs,
obtained by iterated differentiation at zero.
Using this differential formalism, the Taylor formula yields
the weighted sum $\Tay(M)$ of 
all multilinear approximants of a given λ-term $M$,
producing the same total information as \(M\)
\emph{via} normalization.
More precisely, Ehrhard and Regnier's
\enquote{commutation} theorem \cite{EhrhardRegnier08,EhrhardRegnier06}
ensures that the normal form of the Taylor expansion of \(M\)
is the Taylor expansion of the Böhm tree of \(M\):
\begin{equation} \label{eq:intro:commutation}
	\nf(\Tay(M)) = \Tay(\BT(M))
\end{equation}
(and a Böhm tree is uniquely determined by its Taylor expansion).
This approach subsumes the previous one, in the sense that
many results traditionally obtained \emph{via} continuous approximation
enjoy simpler proofs based on linear approximation
\cite{BarbarossaManzonetto20},
and that the continuous approximation theorem itself
can be proved using Taylor expansion.
In addition, it
allows for characterising quantitative properties of programs
(\eg time complexity bounds \cite{deCarvalho.17}),
which is a key benefit of linearity.
This approximation technique
has been fruitfully applied to many languages,
richer than the pure λ-calculus:
nondeterministic \cite{Vaux19}, 
probabilistic \cite{DalLagoLeventis19}, 
extensional \cite{BlondeauPatissierEtAl24},
call-by-value \cite{KerinecEtAl20}, and call-by-push-value 
\cite{EhrhardGuerrieri16,ChouquetTasson20} calculi, 
as well as for Parigot's λμ-calculus \cite{Barbarossa22}.
The interplay between operational properties and Taylor approximations
also suggests a broader notion of approximation of
a computation process \cite{Mazza21,DufourMazza24}.

Another benefit of linear approximation is that it
can approximate not only β-normal\-isation
(the information ultimately produced by a program)
but β-reduction
(the \enquote{information flow} along program execution).
In particular, \cref{eq:intro:commutation} can be refined into
\begin{equation} \label{eq:intro:Lrred-simul-breds}
	M \breds N \Rightarrow \Tay(M) \Lrred \Tay(N),
\end{equation}
where $\Lrred$ denotes the so-called \enquote{resource} reduction
acting linearly on approximants.
As highlighted by our previous work
\cite{CerdaVauxAuclair23,CerdaPhD},
this can even be extended to
\begin{equation} \label{eq:intro:Lrred-simul-bredi}
	M \bred^{\infty} N \Rightarrow \Tay(M) \Lrred \Tay(N)
\end{equation}
if one extends the λ-calculus with infinite λ-terms
and an infinitary closure of the β-reduction,
which is a way to internalise infinite computations and their limits
in the λ-calculus \cite{KennawayEtAl97},
\emph{without changing the target language of Taylor expansion
nor extending its dynamics}.

\medskip

This article is interested in the converse of 
\cref{eq:intro:Lrred-simul-breds,eq:intro:Lrred-simul-bredi}:
is the linear approximation of the λ-calculus conservative?
In other terms, we ask whether every resource reduction
from some $\Tay(M)$ to some $\Tay(N)$ corresponds to a
β-reduction sequence from $M$ to $N$.

In the particular case of \emph{normalisation} of finite \(λ\)-terms, 
the question is
easily solved thanks to the commutation expressed by
\cref{eq:intro:commutation}:
if \(\Tay(N)=\nf(\Tay(M))\) then \(N\) must be the normal form of \(M\),
just because \(N\) is a (finite) λ-term and \(\Tay(N)\) is in normal form,
so that \(N=\BT(M)=\nf(M)\)
(normal λ-terms are precisely those λ-terms that are also Böhm trees).
And if one considers possibly infinite terms and the infinitary version of
normalization, \cref{eq:intro:commutation} (together with the injectivity of
Taylor expansion on Böhm trees) is both a simulation 
\emph{and} a conservativity result:
if \(\Tay(N)=\nf(\Tay(M))\) then \(N=\BT(M)\).

In a sense, 
\cref{eq:intro:Lrred-simul-breds,eq:intro:Lrred-simul-bredi}
thus only generalise to the simulation aspect of \cref{eq:intro:commutation}.
And it turns out that the question of conservativity in the general case of a
reduction \(\Tay(M)\Lrred \Tay(M)\) is quite subtle,
and sensitive to the choice of the source language,
as will become evident along the paper.

\paragraph{Content of the paper.}
We first recall the necessary material:
finite and infinitary λ($\bot$)-calculi, 
the resource λ-calculus and Taylor expansion
(\cref{sec:prelim}).
We then show that the finite β-reduction of finite λ-terms
is conservatively approximated,
\ie the converse of \cref{eq:intro:Lrred-simul-breds} holds 
(\cref{sec:mashup}).
On the contrary, we are able to design a counterexample to conservativity
as soon as we want to approximate infinitary β-reductions.
Defining this λ-term, the \enquote{Accordion} $\acc$,
and proving that it violates the converse of 
\cref{eq:intro:Lrred-simul-bredi},
is the second technical development of this article (\cref{sec:acc}).
However, we do also introduce a \emph{uniform} linear approximation
which still simulates $\bredi$ while enjoying conservativity:
this identifies the \enquote{sub-system} of the resource λ-calculus
that contains exactly the infinitary λ-calculus.
We also discuss how to adapt it to take into account
the so-called $\bot$-reductions that play a key role 
in infinitary λ-calculus, and are needed to reduce a term to its Böhm tree 
(\cref{sec:uniformity}).
Finally, we review our results in a detailed conclusion
featuring several summarising diagrams (\cref{sec:conclusion}).

\medskip

This is an improved and extended version of the article \cite{Cerda.Vau.25}
published in the proceedings of the 42nd~International Symposium on 
Theoretical Aspects of Computer Science (STACS 2025).
It features rearranged and extended presentations
of \cref{sec:acc:acc-proof,sec:uniformity:conservativity},
which contain the proofs of two key results of the paper
(namely \cref{thm:acc,thm:conserv-cohrredi-bredi}).
It also provides additional proofs that were omitted in the
conference version due to space constraints
(in particular \cref{sec:uniformity} has been considerably expanded)
as well as several informal preparatory developments that will hopefully
provide a more palatable exposition of the technical parts of the article.
Finally, \cref{sec:uniformity:beta-bottom,sec:conclusion}
are almost entirely new.

A previous version of this work also appears as Chapter~5 of
the first author's PhD thesis \autocite{CerdaPhD},
but is limited to the qualitative Taylor expansion
(\ie the case where sums of approximants are treated as sets);
extending it to the full, quantitative linear approximation
solves what was presented as Conjecture~5.15 in the thesis.

%% --------------------------------------------------------------------------

\section{Preliminaries} \label{sec:prelim}

In this section, we briefly recall the linear approximation
of the λ-calculus, following its refined presentation in
\cite{CerdaPhD}.
We first recall the definition of the λ($\bot$)-calculus,
as well as its \enquote{001} infinitary extension:
this is the version of the infinitary λ-calculus
that fits the formalism of both continuous and linear approximations
as they are usually presented (\cref{sec:prelim:lcalc}).
Then we present the resource λ-calculus,
\ie a multilinear variant of the λ-calculus
(there are no duplications or erasures of subterms
during the reduction) enjoying strong confluence
and normalisation properties (\cref{sec:prelim:rcalc}).
Finally, the linear approximation relies on the
Taylor expansion operator, that maps a λ-term
to a sum of resource terms,
in a way such that the reduction of λ-terms
is simulated by the reduction of the resource approximants
(\cref{sec:prelim:taylor}).

\subsection{Finite and infinitary λ(\texorpdfstring{$\bot$}{⊥})-calculi} 
\label{sec:prelim:lcalc}

We give a brief presentation of the 001-infinitary λ-calculus
(and of its extension with an \enquote{undefined} term $\bot$
and corresponding rewriting rules, which is a usual construction).
A more detailed exposition and a general account of infinitary λ-calculi
can be found in \cite{CerdaPhD,BarendregtManzonetto22}.

From now on, we fix a countable set $\Vars$ of variables.

\begin{defi} \label{defi:terms}
	Consider the following derivations rules:
	\begin{prooftreeset}
	\begin{prooftree}
		\hypo{ & x \in \Vars }
		\infer1[\Vars]{ & x \in \mathcal{X} }
		\end{prooftree}
	\qquad
		\begin{prooftree}
		\infer0[\bot]{ \bot \in \mathcal X }
		\end{prooftree}
	\qquad
		\begin{prooftree}
		\hypo{x \in \Vars }
		\hypo{ M \in \mathcal{X} }
		\infer2[λ]{ λx.M \in \mathcal{X} }
		\end{prooftree}
	\\[\topsep]
		\begin{prooftree}
		\hypo{ M \in \mathcal{X} }
		\hypo{ N \in \mathcal{X} }
		\infer2[@]{ (M)N \in \mathcal{X}. }
		\end{prooftree}
	\qquad
		\begin{prooftree}
		\hypo{ M \in \mathcal{X} }
		\hypo[doubleprem]{ N \in \mathcal{X} }
		\infer2[@^{001}]{ (M)N \in \mathcal{X} }
		\end{prooftree}
	\end{prooftreeset}
	where in the rule \proofrule{@^{001}}
	the second premise is marked by a double line
	indicating that it is coinductive:
	infinite derivations are allowed provided each infinite branch
	enters infinitely often such a coinductive premise
	(see \autocite{Cerda.Sau.26} for a discussion of this notation).
	\begin{itemize}
	\item The set $\lc$ of (finite) \defemph{λ-terms} is
		the set $\mathcal{X}$ defined by the rules
		\proofrule{\Vars}, \proofrule{λ} and \proofrule{@}.
	\item The set $\lbc$ of (finite) \defemph{λ$\bot$-terms} is
		the set $\mathcal{X}$ defined by the rules
		\proofrule{\Vars}, \proofrule{\bot}, \proofrule{λ} 
		and \proofrule{@}.
	\item The set $\lci$ of \defemph{001-infinitary λ-terms} is
		the set $\mathcal{X}$ defined by the rules
		\proofrule{\Vars}, \proofrule{λ} and \proofrule{@^{001}}.
	\item The set $\lbci$ of \defemph{001-infinitary λ$\bot$-terms} is
		the set $\mathcal{X}$ defined by the rules
		\proofrule{\Vars}, \proofrule{\bot}, \proofrule{λ} 
		and \proofrule{@^{001}}.
	\end{itemize}
	
	In all the following, these sets will be implicitly quotiented:
	\begin{itemize}
	\item as usual by α-equivalence, \ie the equivalence relation
		generated by $λx.M = λy.\subst My$
		(for all term $M$ and variables $x,y$ such that
		$y$ is fresh in $M$, and where
		$\subst M y$ denotes the renaming of $x$ by $y$ in $M$),
		and by lifting to contexts;
	\item by the equivalence relation generated by the identities
		$λx.\bot = (\bot)M = \bot$ for all term $M$,
		and by lifting to contexts.
		(The reason for this additional quotient will become clear after
		\cref{defi:extension}.)
	\end{itemize}
\end{defi}

Notice that we use Krivine's notation for applications
\cite{Krivine90}, \ie 
we parenthesise functions instead of arguments.
We abbreviate the application of a term to successive arguments
\((\cdots((M)N_1)\cdots)N_k\) as \((M)N_1\cdots N_k\),
which is obtained by nesting applications on the left:
this allows to use parentheses more sparingly,
which will be a great relief later on.
By contrast, \((M_1)(M_2)\cdots(M_k)N\) is obtained 
by nesting applications on the right.

Concretely, the rule \proofrule{@^{001}}
means that $\lci$ (resp. $\lbci$) contains 
the infinitary λ-terms (resp. λ$\bot$-terms) whose syntax tree
contains only infinite branches entering infinitely often
the argument side of an application.
A typical example of a term in $\lci$ is
\[\infrapp{x} \eqdef (x)(x)(x)\dots.\]
Observe also that there is an immediate inclusion $\lc \subseteq \lci$.
On the contrary, neither $λx_0.λx_1.λx_2.\dots$
nor $(((\dots)x_2)x_1)x_0$ are allowed in $\lci$.

As the cautious reader may object, it is not obvious at all how to define
the quotient by α-equivalence on the sets $\lci$ and $\lbci$
of infinitary terms;
to do so, we also implicitely restrict these sets to their subsets
of terms having only finitely many free variables,
an innocuous restriction allowing to handle α-equivalence properly 
\cite{Kurz.Pet.Sev.Vri.13,Cerda25}.
This enables us to define capture-avoiding substitution
in the usual way, and we denote by $\subst MN$ the term obtained
by substituting $N$ for $x$ in $M$.

In the following we will equip these sets of terms with
several reduction relations:
some define the dynamics of interest (ordinary β-reduction and
infinitary variants of it, possibly with erasure of unsolvable terms)
while others are introduced as technical, auxiliary notions.
Most of them will come in two flavours, depending on whether
they are solely based on the β-reduction step \((λx.M)N \red \subst MN\),
or if they also include an erasure step \(M\red \bot\).
In order to avoid the tedious repetition of reduction rules,
we will introduce rules parametrized by a reduction symbol \(ξ\),
with only two possible values: \(ξ=β\) or \(ξ=β\bot\).

\begin{defi} \label{defi:bred}
	The relation $\mathord{\bred} \subset \lbci \times \lbci$
	of \defemph{β-reduction} is the relation
	defined by the following base case:
	\begin{prooftreeset}
		\begin{prooftree}
                    \infer0[beta_\xi]{ (λx.M)N \red[\xi] \subst MN }
		\end{prooftree}
	\end{prooftreeset}
	and by the following lifting rules:
	\begin{prooftreeset}
		\begin{prooftree}
		\hypo{ P \red[\xi] P' }
		\infer1[λ_\xi]{ λx.P \red[\xi] λx.P' }
		\end{prooftree}
	\qquad
		\begin{prooftree}
		\hypo{ P \red[\xi] P' }
		\infer1[@l_\xi]{ (P)Q \red[\xi] (P')Q }
		\end{prooftree}
	\qquad
		\begin{prooftree}
		\hypo{ Q \red[\xi] Q' }
		\infer1[@r_\xi]{ (P)Q \red[\xi] (P)Q'. }
		\end{prooftree}
	\end{prooftreeset}
	where the generic symbol $\xi$ is taken to be $β$.
\end{defi}

Recall that a λ$\bot$-term $M$ is said to have 
a \defemph{head normal form} (\textsc{hnf})
whenever there is a reduction
\[	M \breds λx_1.\dots λx_m.(y) M_1 \dots M_n \]
for some variables $x_1, \dots, x_m, y$ and terms $M_1, \dots, M_n$,
and where as usual
$\breds$ denotes the reflexive-transitive closure of $\bred$
(see \cref{defi:head} for a full reminder of the definition
and the associated properties).

\begin{defi} \label{defi:bbotred}
	Consider the following base case\footnote{%
		Usually one also considers two other bases cases,
		namely $λx.\bot \red[\xi] \bot$ and $(\bot)M \red[\xi] \bot$.
		However the sets of terms,
		as we defined them in \cref{defi:terms},
		are already quotiented by the corresponding equalities.
	}:
	\[
                \begin{prooftree}[center]
		\hypo{ M \text{ has no \textsc{hnf}} }
		\infer1[erase_\xi]{ M \red[\xi] \bot }
		\end{prooftree}
                \,
                .
	\]
	
	The relation $\mathord{\bbotred} \subset \lbci \times \lbci$
	of \defemph{β$\bot$-reduction} is defined by base cases
        \proofrule{beta_{β\bot}} and \proofrule{erase_{β\bot}} 
	and by the liftings \proofrule{λ_{β\bot}}, \proofrule{@l_{β\bot}}
	and \proofrule{@r_{β\bot}} from \cref{defi:bred}.
\end{defi}

\begin{defi} \label{defi:bredi}
	The relation $\mathord{\bredi} \subset \lbci \times \lbci$
	(resp. $\bbotredi$)
	of \defemph{001-infinitary β-reduction}
	(resp. β$\bot$-reduction)
	is defined by the rules:
	\begin{prooftreeset}
		\begin{prooftree}
		\hypo{ M \reds[\xi] x }
		\infer1[\Vars_\xi^{001}]{ M \redi[\xi] x }
		\end{prooftree}
	\qquad
		\begin{prooftree}
		\hypo{ M \reds[\xi] \bot }
		\infer1[\bot_\xi^{001}]{ M \redi[\xi] \bot }
		\end{prooftree}
	\qquad
		\begin{prooftree}
		\hypo{ M \reds[\xi] λx.P }
		\hypo{ P \redi[\xi] P' }
		\infer2[λ_\xi^{001}]{ M \redi[\xi] λx.P' }
		\end{prooftree}
	\\[\topsep]
		\begin{prooftree}
		\hypo{ M \reds[\xi] (P)Q }
		\hypo{ P \redi[\xi] P' }
		\hypo[doubleprem]{ Q \redi[\xi] Q' }
		\infer3[@_\xi^{001}]{ M \redi[\xi] (P')Q' }
		\end{prooftree}
%	\qquad
%		\begin{prooftree}
%		\hypo{ & Q \redi Q' }
%		\infer[double]1[\later]{ \later & Q \redi Q' }
%		\end{prooftree}
	\end{prooftreeset}
	where $\xi$ is taken to be $β$ (resp. $β\bot$).
\end{defi}

Infinitary β-reduction can be understood as allowing an infinite number
of β-reduction steps, as long as the β-redexes are fired
inside increasingly nested arguments of applications.
This is formalised in the following result:

\begin{lem}[stratification] \label{lem:stratification}
	Given $M,N \in \lci$, there is a reduction $M \bredi N$ iff
	there exists a sequence of terms $(M_d) \in (\lci)^{\Nats}$
	such that for all $d\in\Nats$,
	\begin{equation*}
		M = M_0 \breds[≥0] M_1 \breds[≥1] M_2 \breds[≥2] \dots
		\breds[≥d-1] M_d \bredi[≥d] N,
	\end{equation*}
	where $\breds[≥d]$ and $\bredi[≥d]$ denote β-reductions
	occurring inside (at least) $d$ nested arguments of applications.
	The result still holds:
	\begin{itemize}
	\item if $\lci$ is replaced with $\lbci$,
	\item if $\lci$ is replaced with $\lbci$
		and β-reductions are replaced with β$\bot$-reductions.
	\end{itemize}
	Formally, for $\xi \in \{β, β\bot\}$,
	\defemph{$\xi$-reduction at minimum depth d}
	is defined by:
	\begin{prooftreeset}
		\begin{prooftree}
			\hypo{ M \red[\xi] M' }
			\infer1[\xi≥0]{ M \red[\xi≥0] M' }
		\end{prooftree}
	\qquad
		\begin{prooftree}
			\hypo{ P \red[\xi\geq d+1] P' }
			\infer1[λ_{\xi\geq d+1}]{ λx.P \red[\xi\geq d+1] λx.P' }
		\end{prooftree}
	\\[\topsep]
		\begin{prooftree}
			\hypo{ P \red[\xi\geq d+1] P' }
			\infer1[@l_{\xi\geq d+1}]{ (P)Q \red[\xi\geq d+1] (P')Q }
		\end{prooftree}
	\qquad
		\begin{prooftree}
			\hypo{ Q \red[\xi\geq d] Q' }
            \infer1[@r_{\xi\geq d+1}]{ (P)Q \red[\xi\geq d+1] (P)Q' }
		\end{prooftree}
	\end{prooftreeset}
	and \defemph{001-infinitary $\xi$-reduction at minimum depth d} is
	defined by:
	\begin{prooftreeset}
		\begin{prooftree}
			\hypo{ M \redi[\xi] M' }
			\infer1[\xi^{001} ≥0]{ M \redi[\xi≥0] M' }
		\end{prooftree}
	\qquad
		\begin{prooftree}
			\infer0[\Vars_{\xi≥d+1}^{001}]{ x \redi[\xi≥d+1] x }
		\end{prooftree}
	\qquad
		\begin{prooftree}
			\infer0[\bot_{\xi≥d+1}^{001}]{ \bot \redi[\xi≥d+1] \bot }
		\end{prooftree}
	\\[\topsep]
		\begin{prooftree}[center]
			\hypo{ P \redi[\xi≥d+1] P' }
			\infer1[λ_{\xi ≥ d+1}^{001}]{ λx.P \redi[\xi≥d+1] λx.P' }
		\end{prooftree}
	\qquad
		\begin{prooftree}[center]
			\hypo{ P \redi[\xi≥d+1] P' }
			\hypo{ Q \redi[\xi≥d] Q' }
			\infer2[@_{\xi≥d+1}^{001}]{ (P)Q \redi[\xi≥d+1] (P')Q'. }
		\end{prooftree}
	\end{prooftreeset}
\end{lem}

A typical (and in fact motivating) example of an infinitary β-reduction
involves the fix-point combinator $\yfp \eqdef λf.(λx.(f)(x)x)λx.(f)(x)x$.
It consists in the reduction $(\yfp)M \bredi \infrapp{M}$
corresponding to the sequence:
\[(\yfp)M \breds[≥0] (M)(\yfp)M \breds[≥1] (M)(M)(\yfp)M \breds[≥2] \dots\]
On the contrary, the infinite reduction sequence 
$\deltadelta \bred \deltadelta \bred \deltadelta \bred \ldots$,
where $\deltadelta \eqdef (λx.(x)x)λx.(x)x$,
does not give rise to a $001$-infinitary reduction
because the redexes are fired at top-level all along the way.
(On the other hand, each finite reduction sequence
$\deltadelta \breds \deltadelta$ induces a reduction
$\deltadelta \bredi \deltadelta$, 
but only because $\bredi$ contains $\breds$;
see \cite{CerdaVauxAuclair23}, Lemma~2.13).

\subsection{The resource λ-calculus} \label{sec:prelim:rcalc}

The resource λ-calculus is the target language of the linear approximation
of the λ-calculus. We recall its construction, and we refer to
\cite{Vaux19,CerdaPhD} for more details.
The main intuition behind this calculus is that arguments become
\emph{finite multisets}, and that $(λx.s)[t_1,\dots,t_n]$
will reduce to a term obtained by substituting \emph{linearly}
one $t_i$ for each occurrence of $x$ in $s$.
The different matchings of the $t_i$'s and the occurrences of $x$
are superposed by a sum operator;
if a wrong number of $t_i$'s is provided,
the term collapses to the empty sum.

Given a set $\mathcal{X}$, we denote by $!\mathcal{X}$
the set of finite multisets of elements of $\mathcal{X}$.
A multiset is denoted by $\ms x = [x_1,\dots,x_n]$,
with its elements in an arbitrary order.
Multiset union is denoted multiplicatively, by $\ms x \cdot \ms y$.
Accordingly, the empty multiset is denoted by~$1$.
We may also write \([x_1^{k_1},\ldots, x_m^{k_m}]\)
to indicate multiplicities:
this is the same as \([x_1]^{k_1}\cdot \ldots\cdot [x_m]^{k_m}\).
%By extension, $\ms x \cdot y$ denotes $\ms x \cdot [y]$.

\begin{defi}
	The set $\rc$ of \defemph{resource terms}
	is defined by the rules:
	\begin{prooftreeset}
	\begin{prooftree}
		\hypo{ & x \in \Vars }
		\infer1[\Vars]{ & x \in \rc }
		\end{prooftree}
	\qquad
		\begin{prooftree}
		\hypo{x \in \Vars }
		\hypo{ s \in \rc }
		\infer2[λ]{ λx.s \in \rc }
		\end{prooftree}
	\qquad
		\begin{prooftree}
		\hypo{ s \in \rc }
		\hypo{ \ms t \in \rc* }
		\infer2[@!]{ (s)\ms t \in \rc }
		\end{prooftree}
	\end{prooftreeset}
	and is implicitely quotiented by α-equivalence.
	Multisets in $\rc*$ are called \defemph{resource monomials}.
	To denote indistinctly $\rc$ or $\rc*$, we write $\rc**$.
\end{defi}

Given a semiring $\SSS$ and a set $\mathcal{X}$, we denote by 
$\SSS^{\mathcal{X}}$ the set of possibly infinite linear combinations
of elements of $\mathcal{X}$ with coefficients in $\SSS$,
considered as formal weighted sums.
Given a sum $\SS \in \SSS^{\mathcal{X}}$, its support $\sumsupp{\SS}$
is the set of all elements of $\mathcal{X}$ bearing a non-null coefficient;
we denote by $\SSS^{(\mathcal{X})}$ the sub-semimodule
of $\SSS^{\mathcal{X}}$ of all sums having a finite support.
The inclusion on sums is the notation defined by writing
$\sum_{x \in \mathcal{X}} a_x \cdot x \suminclusion
\sum_{x \in \mathcal{X}} b_x \cdot x$
whenever for all $x \in \mathcal{X}$
there is a coefficient $a'_x \in \SSS$
such that $a_x + a'_x = b_x$.

We use the following syntactic sugar.
The empty sum $\sum_{x \in \mathcal{X}} 0\cdot x$ is denoted by $0$.
The one-element sum $\sum_{x \in \mathcal{X}} \delta_{x,y}\cdot x$
is assimilated to $y$, yielding an inclusion 
$\mathcal{X} \subseteq \SSS^{\mathcal{X}}$.
Sums can be summed, \ie 
$\sum_{x \in \mathcal{X}} a_x \cdot x
+\sum_{x \in \mathcal{X}} b_x \cdot x
=\sum_{x \in \mathcal{X}} (a_x+b_x)\cdot x$.

In practice, we will work with the sets $\irsums$ and $\irsums*$
of sums of resource terms and monomials. In this setting,
it is convenient to extend by linearity
all the constructors of the calculus
to sums of resource terms, \ie 
\begin{equation} \label{eq:prelim:notation-linearity}
\begin{aligned} 
	&\textstyle λx.\left(\sum\limits_{i \in I} a_i \cdot s_i\right) 
		\eqdef \sum\limits_{i \in I} a_i \cdot λx.s_i, \\
	&\textstyle \left(\sum\limits_{i \in I} a_i \cdot s_i\right)
		\sum\limits_{j \in J} b_j \cdot \ms t_j
		\eqdef \sum\limits_{i \in I} \sum\limits_{j \in J} 
		a_i b_j \cdot (s_i)\ms t_j, \\
        &\textstyle \left[ \sum\limits_{i \in I} a_i \cdot s_i \right] \cdot
		\sum\limits_{j \in J} b_j \cdot \ms t 
		\eqdef \sum\limits_{i\in I} \sum\limits_{j\in J} 
                a_i b_j\, \cdot\, [s_i]\cdot \ms t_j\,.
\end{aligned}
\end{equation}

\begin{defi}
	For all $u \in \rc**$, $\ms t = [t_1,\dots,t_n] \in \rc*$
	and $x \in \Vars$, the \defemph{multilinear substitution}
	of $x$ by $\ms t$ in $u$ is the finite sum
	$\rsubst st \in \frsums**$ defined by
	\[	\rsubst st \eqdef \left\lbrace \begin{array}{cl}
		\displaystyle\sum _{ \sigma \in \perm(n) }
			u[ t_{\sigma(1)}/x_1, \dots, t_{\sigma(n)}/x_n ]
			& \text{if $x$ occurs $n$ times in $u$} \\
		0 & \text{otherwise,}
		\end{array} \right. \]
	where $x_1,\dots,x_n$ is an arbitrary enumeration of the occurrences
	of $x$ in $u$, and $u[ t_{\sigma(1)}/x_1, \dots]$
	denotes the result of the (capture-avoiding) substitution
	of each $x_i$ by the corresponding $t_{\sigma(i)}$.
\end{defi}

\begin{defi}
	The relation $\mathord{\rred} \subset \frsums** \times \frsums**$
	of \defemph{resource β-reduction} is defined
	using the auxiliary relation 
	$\mathord{\Orred} \subset \rc** \times \frsums**$
	generated by the rules
	\begin{prooftreeset}
		\begin{prooftree}
			\infer0[beta_{\resource}]{
				\rapp{λx.s}t \Orred \rsubst st }
		\end{prooftree}
	\qquad
		\begin{prooftree}
			\hypo{ s \Orred S' }
			\infer1[λ_{\resource}]{ λx.s \Orred λx.S' }
		\end{prooftree}
	\qquad
		\begin{prooftree}
			\hypo{ s \Orred S' }
			\infer1[@l_{\resource}]{ \rapp st \Orred \rapp{S'}t }
		\end{prooftree}
	\\[\topsep]
		\begin{prooftree}
			\hypo{ \ms t \Orred \ms T' }
			\infer1[@r_{\resource}]{ \rapp st \Orred \rapp s T' }
		\end{prooftree}
	\qquad
		\begin{prooftree}
			\hypo{ s \Orred S' }
                        \infer1[!_{\resource}]{ [s] \cdot \ms t \Orred 
                        {[S']} \cdot \ms t }
		\end{prooftree}
	\end{prooftreeset}
	as well as the lifting rule
	\begin{prooftreeset}
		\begin{prooftree}
		\hypo{ u_1 \Orred U'_1 }
		\hypo{ \forall i≥2,\ u_i \Orredr U'_i }
		\infer2[\Sigma_{\resource}]{ \sum_{i=1}^n u_i \rred \sum_{i=1}^n 
		U'_i }
		\end{prooftree}
	\end{prooftreeset}
	where $\Orredr$ is the reflexive closure of $\Orred$.
\end{defi}

From now on, we fix a semiring $\SSS$.
We consider $\Nats$ as a subset of $\SSS$ through the map
$n \mapsto 1+\ldots+1$ (notice however that it might not be an injection), 
and we suppose that $\SSS$ \enquote{has fractions},
\ie for all non-null $n \in \Nats$ there is some $\frac 1n \in \SSS$
such that $n \times \frac 1n = 1$.
This is the case of the semirings $\mathbb{Q}_+$ and $\mathbb{R}_+$
of non-negative rational (resp.~real) numbers,
but also of the semiring $\mathbb{B}$ of boolean values
(equipped with the logical \enquote{or} and \enquote{and} operations).

\begin{defi}
	Given a set $\mathcal{X}$ and a semiring $\SSS$,
	a family of sums $(\SS_i)_{i \in I} \in (\SSS^{\mathcal{X}})^I$
	is \defemph{summable} when each $x \in \mathcal{X}$
	bears a non-null coefficient in finitely many of the $\SS_i$.
	If this is the case then $\sum_{i \in I} \SS_i$ is a well-defined sum.
\end{defi}

\begin{defi}
	The relation $\mathord{\Lrred} \subset \irsums** \times \irsums**$ 
	of \defemph{pointwise resource reduction} is defined by saying that
	there is a reduction $\UU \Lrred \VV$
	whenever there are summable families
	$(u_i)_{i\in I} \in (\rc**)^I$ and
	$(V_i)_{i\in I} \in (\frsums**)^I$
	such that
	\[	\UU = \sum_{i \in I} a_i \cdot u_i, \quad
		\VV = \sum_{i \in I} a_i \cdot V_i
		\quad \text{and} \quad
		\forall i\in I,\ u_i \rreds V_i. \]
\end{defi}

Notice that whereas $\rred$ reduces finite sums
with integer coefficients,
$\Lrred$ reduces arbitrary sums with arbitrary coefficients.

\subsection{Linear approximation and the conservativity problems}
\label{sec:prelim:taylor}

We recall the definition of the Taylor expansion of λ-terms,
and the approximation theorems it enjoys.
Again, a detailed presentation can be found in \cite{Vaux19},
and in \cite{CerdaPhD} for the adaption to infinitary λ-calculi.
In the latter setting, we shall start with the following
unusual definition.

\begin{defi} \label{defi:Tay}
	The \defemph{Taylor expansion} is the map $\Tay : \lbci \to \irsums$
	defined by \[ \Tay(M) \eqdef \sum_{s \in \rc} \Tay(M,s) \cdot s, \]
	where the coefficient $\Tay(M,s)$ is defined 
	by induction on $s \in \rc$ as follows:
	\begin{align*}
		\Tay(x,x) & \eqdef 1 \\
		\Tay(λx.P, λx.s) & \eqdef \Tay(P,s) \\
		\Tay((P)Q, (s)\ms t) & \eqdef \Tay(P,s) \times \Tay^!(Q, \ms t) \\
	\shortintertext{where, for pairwise distinct $t_i$'s, we denote:}
		\Tay^!(Q, [t_1^{k_1}, \ldots, t_m^{k_m}])
			& \eqdef \textstyle\prod\limits_{i=1}^{m}
			\tfrac{ \Tay(Q,t_i)^{k_i} }{ k_i! }, \\
	\shortintertext{and in all other cases
		(\ie whenever $M$ and $s$ do not have the same shape):}
		\Tay(M,s) & \eqdef 0.
	\end{align*}
\end{defi}

Let us stress a crucial observation: whenever $s \in \sumsupp{\Tay(M)}$,
the value of $\Tay(M,s)$ does not depend on $M$,
hence \(\Tay(M)\) is uniquely determined by its 
support~\cite{EhrhardRegnier08}.

Using the notation from \cref{eq:prelim:notation-linearity}, 
we obtain the following description
of the Taylor expansion.
This is usually how the definition is presented
for finite λ-terms,
but since it is not a valid coinductive definition
we had to provide \cref{defi:Tay} in the infinitary setting.

\begin{lem}[{\cite{CerdaPhD}}, Corollary 4.7]
	For all variables $x \in \Vars$ and terms $P,Q \in \lbci$,
	\[	\Tay(x) = x \qquad 
		\Tay(\bot) = 0 \qquad
		\Tay(λx.P) = λx.\Tay(P) \qquad
		\Tay((P)Q) = (\Tay(P)) \Tay(Q)^!, \]
	where the operation of \defemph{promotion} is defined
	for all $\SS \in \irsums$ by
	$	\SS^! \eqdef \sum\limits_{n \in \Nats} \frac 1{n!} \cdot [ \SS ]^n$.
\end{lem}

We defined a map $\Tay$ taking λ-terms to weighted sums of approximants.
This induces an approximation of the λ-calculus, thanks to
the following theorems expressing the fact that
the reduction of the approximants can simulate the reduction
of the approximated term.

\begin{simulthm}[{\cite{Vaux19}}, Lemma~7.6]
\label{thm:Lrred-simul-breds}
	For all $M,N \in \lc$, if $M \breds N$ then $\Tay(M) \Lrred \Tay(N)$.
\end{simulthm}

\begin{simulthm}[{\cite{CerdaVauxAuclair23}}, Theorem 4.21]
\label{thm:Lrred-simul-bredi}
	For all $M,N \in \lci$, if $M \bredi N$ then $\Tay(M) \Lrred \Tay(N)$.
\end{simulthm}

This second theorem can in fact be strengthened:

\begin{simulcor}[{\cite{CerdaVauxAuclair23}}, Corollary 5.13]
\label{cor:Lrred-simul-bbotredi}
	For all $M,N \in \lbci$, if $M \bbotredi N$
	then $\Tay(M) \Lrred \Tay(N)$.
\end{simulcor}
Note that,
in the given reference~\cite{CerdaVauxAuclair23},
\cref{thm:Lrred-simul-bredi,cor:Lrred-simul-bbotredi}
were in fact presented and proved only for the qualitative Taylor expansion
(\ie when $\SSS$ is the semiring of booleans).
The proof of the quantitative version (for any semiring)
can be found in \cite[Theorem 4.56]{CerdaPhD}.

In particular, the \cref{cor:Lrred-simul-bbotredi} encompasses the
\enquote{Commutation theorem} \cite{EhrhardRegnier08,EhrhardRegnier06},
which is usually presented as the cornerstone of the linear approximation
of the λ-calculus:
the normal form of $\Tay(M)$ is equal to 
the Taylor expansion of the Böhm tree of $M$
(which is a notion of infinitary β-normal form of $M$),
\ie normalisation commutes with approximation.

\begin{defi} \label{defi:extension}
	Let $(A, \red[A])$ and $(B, \red[B])$
	be two reduction systems.
	The latter is an \defemph{extension} of the former if:
	\begin{enumerate}
	\item there is an injection $i : A \hookrightarrow B$,
	\item $\mathord{\red[A]}$ \defemph{simulates} $\mathord{\red[B]}$
		through $i$, \ie 
		$\forall a,a' \in A$, if $a \red[A] a'$
		then $i(a) \red[B] i(a')$.
	\end{enumerate}
	This extension is said to be \defemph{conservative}
	if $\forall a,a' \in A$, if $i(a) \red[B] i(a')$ then $a \red[A] a'$.
\end{defi}

Notice that our definition of a conservative extension
varies from the one chosen by the \emph{Terese} \cite[§~1.3.21]{Terese},
where the conservativity of $\red[B]$ \wrt $\red[A]$ is defined
as a property of the conversions $=_A$ and $=_B$ they generate.
We prefer to distinguish between a conservative extension of a reduction
(\enquote{in the image of the small world,
the big reduction relates the same people
than the small reduction did})
and a conservative extension of the corresponding conversion.

An important observation is that the map $\Tay : \lbci \to \irsums$ 
is injective \cite[Lemma~5.18]{CerdaVauxAuclair23};
in fact, it is only to ensure this that we quotiented λ$\bot$-terms
by $λx.\bot = \bot$ and $(\bot)M = \bot$ in \cref{defi:terms}.
As a consequence, \cref{thm:Lrred-simul-breds,thm:Lrred-simul-bredi}
can be reformulated with the terminology of \cref{defi:extension}:
\begin{itemize}
\item \cref{thm:Lrred-simul-breds} tells that
	$(\irsums, \Lrred)$ simulates $(\lc, \breds)$,
\item \cref{thm:Lrred-simul-bredi} tells that
	$(\irsums, \Lrred)$ simulates $(\lci, \bredi)$,
\end{itemize}
which leads us to the problems we tackle in this article.

\begin{prob} \label{problem:finite}
	Is $(\irsums, \Lrred)$ conservative \wrt $(\lc, \breds)$?
\end{prob}

\begin{prob} \label{problem:infinitary}
	Is $(\irsums, \Lrred)$ conservative \wrt $(\lci, \bredi)$?
\end{prob}

\Cref{sec:mashup,sec:acc} are devoted to giving
(respectively positive and negative) answers
to these problems.
Let us just tell the reader that this may be the right moment for them
to have a first look at \cref{sec:conclusion}, 
where we provide an overview of our results,
before coming back to the following technical developments.

%*******************************************************************************

\section{Conservativity \texorpdfstring{\wrt}{wrt.}
the finite {\upshape λ}-calculus}
\label{sec:mashup}

In this first section, we prove the following result
provinding a positive answer to \cref{problem:finite}:

\begin{conservthm}
\label{thm:conserv-Lrred-breds}
	For all $M, N \in \lc$, if $\Tay(M) \Lrred \Tay(N)$
	then $M \breds N$.
\end{conservthm}

To do so, we adapt a proof technique by 
Kerinec and the second author~\cite{KerinecVauxAuclair23},
who used it to prove that the algebraic λ-calculus
is a conservative extension of the usual λ-calculus.
Their proof relies on a relation $\mashup$, 
called \enquote{mashup} of β-reductions,
relating λ-terms (from the \enquote{small world})
to their algebraic reducts (in the \enquote{big world}).
In our setting, $M \mashup s$ when $s$ is 
an approximant of a reduct of $M$.

\begin{defi} \label{defi:mashup}
	The \defemph{mashup} relation $\mathord{\mashup} \subset \lc \times \rc$
	is defined by the following rules:
	\begin{prooftreeset}
		\begin{prooftree}
			\hypo{ M \breds x }
			\infer1[\mathord{\mashup}\Vars]{ M \mashup x }
		\end{prooftree}
	\qquad
		\begin{prooftree}
			\hypo{ M \breds λ x.P }
			\hypo{ P \mashup s }
			\infer2[\mathord{\mashup}λ]{ M \mashup λx.s }
		\end{prooftree}
	\\[\topsep]
		\begin{prooftree}
			\hypo{ M \breds (P)Q }
			\hypo{ P \mashup s }
			\hypo{ Q \mashup \ms t }
			\infer3[\mathord{\mashup}@]{ M \mashup \rapp st}
		\end{prooftree}
	\qquad
		\begin{prooftree}
			\hypo{ M \mashup t_1 }
			\hypo{\dots}
			\hypo{ M \mashup t_n }
            \infer3[\mathord{\mashup}!]{ M \mashup {[t_1,\dots,t_n]} }
		\end{prooftree}
	\end{prooftreeset}
	It is extended to $\lc \times \irsums$
	by the following rule:
	\begin{prooftreeset}
	\begin{prooftree}
		\hypo{ \forall i \in I,\ M \mashup s_i }
		\infer1[\mathord{\mashup}\Sigma]
			{ M \mashup \sum_{i \in I} a_i \cdot s_i }
	\end{prooftree}
	\end{prooftreeset}
	for any index set $I$ and coefficients $a_i \in \SSS$
	such that the sum exists.
\end{defi}

\begin{lem} \label{lem:mashup-1refl}
	For all $M \in \lc$, $M \mashup \Tay(M)$.
\end{lem}

	\begin{proof}
	Take any $s \in \sumsupp{\Tay(M)}$. 
	By an immediate induction on $s$,
	$M \mashup 	s$ 
	follows from the rules of \cref{defi:mashup}
	(where all the assumptions $\breds$ are just taken to be equalities).
	\end{proof}

\begin{lem} \label{lem:mashup-2flb}
	For all $M, N \in \lc$ and $\SS \in \irsums$,
	if $M \breds N$ and $N \mashup \SS$ then $M \mashup \SS$.
\end{lem}

	\begin{proof}
	Take any $s \in \sumsupp{\SS}$, then $N \mashup s$.
	By an immediate induction on $s$, $M \mashup s$
	follows from the rules of \cref{defi:mashup}
	(where the assumptions $M \breds \dots$ follow from 
	the corresponding $M \breds N \breds \dots$).
	\end{proof}

\begin{lem} \label{lem:mashup-3subst}
	For all $M, N \in \lc$, $x \in \Vars$, $s \in \rc$
	and $\ms t \in \rc*$,
	if $M \mashup s$ and $N \mashup \ms t$
	then $\forall s' \in \sumsupp{\rsubst st},\ \subst MN \mashup s'$.
\end{lem}

	\begin{proof}
	Assume $M$ and $N$ are given and show the following equivalent result
	by induction on $s$:
	if $M \mashup s$ then for all $\ms t$ such that $N \mashup \ms t$
	and for all $s'\in \sumsupp{\rsubst st}$, $\subst MN \mashup s'$.
	
	\begin{itemize}

	\item If $s = x$, then $\ms t=[t_1]$ and $s'= t_1$.
		Since $M \mashup x$ and  $N \mashup{[t_1]}$, we have $M \breds x$
		and we obtain $\subst M N \breds N \mashup t_1 = s'$.

	\item If $s = y\not=x$, then $\ms t=1$ and $s'=y$.
		Since $M \mashup y$, we have $M \breds y$
		and we obtain $\subst M N \breds y$ hence $\subst M N \mashup y$.

	\item If $s = λx.u$,
		then $s' \in \sumsupp{λx.\rsubst ut}$,
		that is $s' = λx.u'$ for some $u' \in \sumsupp{\rsubst ut}$.
		Since $M \mashup λx.u$,
		there is some $M \breds λx.P$ with $P \mashup u$.
		By induction hypothesis, $\subst PN \mashup u'$.
		Hence $\subst M N \breds λx. \subst P N$
		and $\subst P N \mashup u'$, so $\subst M N \mashup λx.u'$.
	
	\item If $s = \rapp uv$ with $\ms v=[v_1,\dots,v_n]$,
		then $s' = \rapp{u'}v'$ with $u' \in \sumsupp{\rsubst*{u}{\ms t_0}}$,
        $\ms v'=[v'_1,\dots,v'_n]$ and 
        $v'_i \in \sumsupp{\rsubst*{v_i}{\ms t_i}}$
        for $i \in \{1,\dots,n\}$,
        so that $\ms t = \ms t_0 \cdot \ms t_1 \cdot \ldots \cdot \ms t_n$.
		Since $M \mashup \rapp uv$, 
		there is some $M \breds (P)Q$ with $P \mashup u$ and
		$Q \mashup \ms v$.
		Since $N\mashup\ms t$, we also have $N\mashup\ms t_i$
		for each $i \in \{0,\dots,n\}$.
        By induction hypothesis, we obtain $\subst P N \mashup u'$
        and $\subst Q N \mashup v'_i$ for each $i \in [1,n]$.
		Therefore $\subst M N \breds \left( \subst P N \right) \subst Q N$
		with $\subst P N \mashup u'$ and $\subst Q N \mashup \ms v'$,
		so finally $\subst M N \mashup \rapp{u'}v'$. \qedhere
	\end{itemize}
	\end{proof}

\begin{lem} \label{lem:mashup-4flr}
	For all $M \in \lc$ and $\SS, \TT \in \irsums$,
	if $M \mashup \SS$ and $\SS \Lrred \TT$
	then $M \mashup \TT$.
\end{lem}

	\begin{proof}
	Let us first show that for all $M \in \lc$ and $s \in \rc$
	and $T \in \frsums$, if
	$M \mashup s \Orred T$ then
	$\forall t \in \sumsupp{T}$, $M \mashup t$.
	We do so by induction on $s \Orred T$. 
	When $s = \rapp{λx.u}v$ is a redex,
	there exists a derivation:
	\begin{prooftree*}
		\hypo{ M \breds (P)Q }
		\hypo{ P \breds λx.P' }
		\hypo{ P' \mashup u }
		\infer2[\mathord{\mashup}λ]{ P \mashup λx.u }
		\hypo{ Q \mashup \ms v }
		\infer3[\mathord{\mashup}@]{ M \mashup \rapp{λx.u}v }
	\end{prooftree*}
	By \cref{lem:mashup-3subst} with $P' \mashup u$,
	$Q \mashup \ms v$,
	for all $t \in \sumsupp{\rsubst uv}$,
	we obtain $\subst{P'}{Q} \mashup t$.
	Finally, since $M \breds (λx.P')Q \bred \subst{P'}{Q}$,
	we concude by \cref{lem:mashup-2flb}.
	The other cases of the induction follow immediately
	by lifting to the context.
	
	As a consequence, we can easily deduce the following steps:
	\begin{itemize}
	\item if $M \mashup s \Orred T$ then $M \mashup T$,
		for all $M \in \lc$, $s \in \rc$ and $T \in \frsums$,
	\item if $M \mashup S \rred T$ then $M \mashup T$,
		for all $M \in \lc$ and $S,T \in \frsums$,
	\item if $M \mashup S \rreds T$ then $M \mashup T$,
		for all $M \in \lc$ and $S,T \in \frsums$,
	\end{itemize}
	which leads to the result.
	\end{proof}

Before we state the last lemma of the proof,
recall that there is a
canonical injection $\lctorc- : \lc \to \rc$ defined by:
\[
	\lctorc{x}		\eqdef x							\qquad
	\lctorc{λx.P}	\eqdef λx.\lctorc P					\qquad 
	\lctorc{(P)Q}	\eqdef \rapp*{\lctorc P}{[\lctorc Q]}
\]
and such that for all $N \in \lc$, $\lctorc N \in \sumsupp{\Tay(N)}$.

\begin{lem} \label{lem:mashup-5tay}
	For all $M, N \in \lc$, if $M \mashup \Tay(N)$ then $M \breds N$.
\end{lem}

	\begin{proof}
	If $M \mashup \Tay(N)$, then in particular $M \mashup \lctorc N$.
	We proceed by induction on $N$:
	
	\begin{itemize}
	\item If $N = x$, then $M \mashup x$ so $M \breds x$ by definition.
	\item If $N = λx.P'$, then $M \mashup λx.\lctorc{P'}$,
		\ie there is a $P \in \lc$ such that $M \breds λx.P$
		and $P \mashup \lctorc{P'}$.
		By induction, $P \breds P'$, thus $M \breds λx.P' = N$.
	\item If $N = (P')Q'$, then 
		$M \mashup \rapp*{\lctorc{P'}}{[\lctorc{Q'}]}$
		\ie there are $P, Q \in \lc$ such that $M \breds (P)Q$,
		$P \mashup \lctorc{P'}$ and $Q \mashup{} [\lctorc{Q'}]$.
		By induction, $P \breds P'$ and $Q \breds Q'$,
		thus $M \breds (P')Q' = N$. \qedhere
	\end{itemize}
	\end{proof}

\begin{proof}[Proof of \cref{thm:conserv-Lrred-breds}]
	Suppose that $\Tay(M) \Lrred \Tay(N)$.
	By \cref{lem:mashup-1refl} we obtain $M \mashup \Tay(M)$,
	hence by \cref{lem:mashup-4flr} $M \mashup \Tay(N)$.
	We can conlude with \cref{lem:mashup-5tay}.
\end{proof}

%*******************************************************************************

\section{Non-conservativity \texorpdfstring{\wrt}{wrt.} 
the infinitary {\upshape λ}-calculus} 
\label{sec:acc}

The previous theorem relied on the excellent properties
of the Taylor expansion of finite λ-terms:
a single (well-chosen) term $\lctorc M \in \sumsupp{\Tay(M)}$
is enough to characterise $M$,
and a single (again, well-chosen) sequence of resource reducts of some
$s \in \sumsupp{\Tay(M)}$ suffices to characterise any sequence $M \breds N$.
These properties are not true any more when considering more complicated 
settings, like the 001-infinitary λ-calculus.
This does not only make the \enquote{mashup} proof technique fail,
but also enables us to give a negative answer to \cref{problem:infinitary}.

\subsection{Failure of the \enquote{mashup} technique}
%-----------------------------------------------------

Let us first describe where we hit an obstacle if we try to reproduce
the proof we have given in the finite setting,
which will make clearer the way we later build a counterexample.

First, it is not obvious what the mashup relation should be:
we could just use the relation $\mashup$ defined on
$\lci \times \rc$ by the same set of rules as in 
\cref{defi:mashup},
or define an infinitary mashup $\mashup[001]$ by the rules
\begin{prooftreeset}
	\begin{prooftree}
		\hypo{ M \bredi x }
		\infer1[\mathord{\mashup[001]}\Vars]{ M \mashup[001] x }
	\end{prooftree}
\qquad
	\begin{prooftree}
		\hypo{ M \bredi λ x.P }
		\hypo{ P \mashup[001] s }
		\infer2[\mathord{\mashup[001]}λ]{ M \mashup[001] λ x.s }
	\end{prooftree}
\\[\topsep]
	\begin{prooftree}
		\hypo{ M \bredi (P)Q }
		\hypo{ P \mashup[001] s }
		\hypo{ Q \mashup[001] \ms t }
		\infer3[\mathord{\mashup[001]}@]{ M \mashup[001] \rapp st}
	\end{prooftree}
\qquad
	\begin{prooftree}
		\hypo{ M \mashup[001] t_1 }
		\hypo{\dots}
		\hypo{ M \mashup[001] t_n }
		\infer3[\mathord{\mashup[001]}!]{ M \mashup[001] [t_1,\dots,t_n] }
	\end{prooftree}
\end{prooftreeset}
and extend it to $\irsums$ accordingly.
In fact, this happens to define the same relation.

\begin{lem} \label{lem:mashup-001-is-mashup}
	For all $M \in \lci$ and $s \in \rc$,
	$M \mashup[001] s$ iff $M \mashup s$.
\end{lem}
	
	\begin{proof}
	The inclusion
	$\mathord{\mashup} \subseteq \mathord{\mashup[001]}$ is immediate.
	Let us show the converse.
	First, observe that the proof of \cref{lem:mashup-2flb}
	can be easily extended in order to show that
	for all $M, N \in \lci$ and $s \in \rc$,
	if $M \bredi N \mashup[001] s$ then $M \mashup[001] s$.
	Then we proceed by induction on $s$.
	\begin{itemize}
	\item If $M \mashup[001] x$, then $M \bredi x$, \ie $M \breds x$,
		and finally $M \mashup x$.
	\item If $M \mashup[001] λ x.u$, then there is a derivation:
		\begin{prooftree*}
			\hypo{ M \breds λ x.P }
			\hypo{ P \bredi P' }
			\infer2[λ_β^{001}]{ M \bredi λ x.P' }
			\hypo{ P' \mashup[001] u }
			\infer2[\mathord{\mashup[001]}λ]{ M \mashup[001] λ x.u }
		\end{prooftree*}
		Since $P \bredi P' \mashup[001] u$,
		we have $P \mashup[001] u$,
		and by induction on $u$ we obtain $P \mashup u$.
		With $M \breds λ x.P$, this yields $M \mashup λ x.u$.
	\item The case of $M \mashup[001] \rapp uv$ is similar. \qedhere
	\end{itemize}
	\end{proof}

As a consequence, \crefrange{lem:mashup-1refl}{lem:mashup-4flr}
can be easily extended to $\bredi$ and $\mashup[001]$. 
We have already explained how the proof of such an extension can be done for
\cref{lem:mashup-2flb};
for the other ones, one just needs to observe that
the proofs are all by induction on 
resource terms or on some inductively defined relation,
hence replacing $\breds$ with $\bredi$
does not change anything
(and neither does replacing $\mashup$ with $\mashup[001]$,
thanks to \cref{lem:mashup-001-is-mashup}).

The failure of the infinitary \enquote{mashup} proof occurs
in the extension of \cref{lem:mashup-5tay}.
Indeed, this proof crucially relies on the existence of an injection
$\lctorc- : \lc \to \rc$,
whereas for $\lci$ there is only the counterpart
$\lctorc{-}[-] : \lci \times \Nats \to \rc$
defined by
\begin{align*}
	\lctorc{ x }[d]				& \eqdef x &
	\lctorc{ (P)Q }[0]			& \eqdef \rapp*{ \lctorc P [0] }1 \\
	\lctorc{ λ x.P }[d]	& \eqdef λ x. \lctorc P [d] &
	\lctorc{ (P)Q }[d+1]		& \eqdef 
		\rapp*{ \lctorc P [d+1] }{\left[ \lctorc Q [d] \right]}.
\end{align*}
Now, if we suppose that $M \mashup \Tay(N)$
and we want to show that $M \bredi N$,
we cannot rely any more on the fact that $M \mashup \lctorc N$,
but only on the fact that
$\forall d\in\Nats,\ M \mashup \lctorc N [d]$.
This makes the induction fail.
For instance, for the case where
$N$ is an abstraction $λx.P'$,
we obtain a $d$-indexed sequence of derivations
\[\begin{prooftree}
	\hypo{ M \breds λ x.P_d }
	\hypo{ P_d \mashup \lctorc{P'}[d] }
	\infer2[\mathord{\mashup}λ]
		{ M \mashup \lctorc N [d] = \lctorc{λ x.P'}[d] }
\end{prooftree}\]
but nothing tells us that the terms $P_d$ and reductions
$M \breds λ x.P_d$ are coherent!
This failure is what enables us to design a counterexample.

\subsection{The Accordion}
%-------------------------
\label{sec:acc:acc-definition}

In this section, we define 001-infinitary λ-terms $\acc$ and $\acc*$
and show that they form a counterexample not only to the 
001-infinitary counterpart of \cref{lem:mashup-5tay},
but also to the conservativity property in the infinitary setting.

As the reader will see below (in \cref{defi:acc}),
the definition of the λ-term $\acc$ turns out to be quite convoluted;
as a consequence, the proof that it is actually a counterexample,
although unsurprising, is boringly technical.
To illustrate the intuitions that led to the design of the term \(\acc\),
let us start with a more naive attempt:
although this will ultimately fail,
it will give the main ideas that we leveraged,
and will also explain why the technical convolutions of the actual
counterexample were introduced.

Let us first introduce some notations.

\begin{nota} \label{nota:before-acc}
	We denote as follows the usual representation of 
	the Church encodings of integers, the successor function, and booleans,
	as well as an \enquote{applicator} $\appterm-$:
	\begin{gather*}
		\churchn \eqdef λ f. λ x. (f)^n x					
		\qquad 
		\churchsucc \eqdef λ n.λ f.λ x. (n) \, f \, (f)x	\\
		\ttterm \eqdef λ x. λ y. x							
		\qquad
		\ffterm \eqdef λ x. λ y. y							
		\qquad
		\appterm M \eqdef λ b. (b) M.
	\end{gather*}
\end{nota}

Now let us try to construct a counterexample to conservativity.
We thus want:
\begin{itemize}
\item a term $\acc$, possibly finite,
\item a term $\acc*$, necessarily infinite
	(otherwise \cref{thm:conserv-Lrred-breds} applies),
	which we will try to build of the following shape,
	for finite terms $B_n$:
	\[
	\begin{tikzpicture}[
	baseline, 
	every node/.style={anchor=base}, 
	xscale=1, yscale=0.8
	]
		\path	(10,0)	node (5a0)	{ $@$ }
		+		(-.4,-1)	node (5t0)	{ $B_0$ }
		++		(.4	,-1)	node (5a1)	{ $@$ }
		+		(-.4,-1)	node (5t1)	{ $B_1$ }
		++		(.4	,-1)	node (5a2)	{ $@$ }
		+		(-.4,-1)	node (5t2)	{ $B_2$ }
		++		(.4	,-1)	node (5an)	{};
		\draw 			(5a0) -- (5t0);
		\draw 			(5a0) -- (5a1);
		\draw 			(5a1) -- (5t1);
		\draw 			(5a1) -- (5a2);
		\draw 			(5a2) -- (5t2);
		\draw[dotted] 	(5a2) -- (5an);
	\end{tikzpicture}
	\]
\end{itemize}
such that the β-reduction of $\acc$ produces
approximations of $\acc*$, as accurate as desired,
\ie there are terms $C_n$ such that
\[	\acc \breds (B_0)C_0 \breds (B_0)(B_1)C_1 \breds 
	(B_0)(B_1)(B_2)C_2 \breds \dots \]
(in particular, the applicative depth of the first difference between 
\((B_0)(B_1)\cdots(B_n)C_n\) and \(\acc*\) tends to infinity)
\emph{but} such that this sequence of β-reductions does not give rise to
a valid infinitary β-reduction $\acc \bredi \acc*$,
\ie in each reduction
$(B_0)\dots(B_n)C_n \breds (B_0)\dots(B_{n+1})C_{n+1}$
one reduction step should occur at a globally bounded depth.
The precise way we want to achieve this is by decomposing each
of the above reductions into
\begin{equation} \label{eq:acc:towards-acc}
	(B_0)\dots(B_n)C_n \breds (\lamP)\,\churchn[n+1]
	\breds (B_0)\dots(B_{n+1})C_{n+1}
\end{equation}
for a given term $\lamP$, so that the second part of 
\cref{eq:acc:towards-acc} 
starts with a reduction occurring at depth~0,
and to define $\acc \eqdef (\lamP)\churchn[0]$.

For the sake of naivety, let us observe that the first part of
\cref{eq:acc:towards-acc} can be achieved by defining
$B_n \eqdef \idterm$ (the identity λ-term $λx.x$)
and $C_n \eqdef (\lamP)\,\churchn[n+1]$.
As for the second part of \cref{eq:acc:towards-acc}, we can define:
\[	\lamP \eqdef (\yfp)λp.λn.(n) \, \idterm \, (p)(\churchsucc)n \]
so that, thanks to the fixed-point combinator,
$(\lamP)\churchn[n] 
\breds (\churchn[n]) \, \idterm \, (\lamP)(\churchsucc)\churchn[n]
\breds (\idterm)^n (\lamP)\, \churchn[n+1]$.
As a result, we obtain:
\[	\acc = (\lamP) \churchn[0]
	\breds (\lamP) \churchn[1]
	\breds (\idterm) (\lamP) \churchn[2]
	\breds (\lamP) \churchn[2]
	\breds (\idterm) (\idterm) (\lamP) \churchn[3]
	\breds (\lamP) \churchn[3]
	\breds \dots \]
which, as expected, does not correspond to a valid
infinitary β-reduction.

However this counterexample does not work, because we do not only want that
\emph{one} sequence of β-reductions starting from $\acc$
and converging to $\acc*$ cannot be turned into a valid
infinitary β-reduction,
but that \emph{all} such sequences enjoy this (lack of) property!
And in the naive case we just presented,
we can in fact reduce $\acc$ as follows:
\begin{equation} \label{eq:acc:bad-counterexample}
	\acc = \fired{(\lamP) \churchn[0]}
	\breds \fired{(\lamP) \churchn[1]}
	\breds (\idterm) \fired{(\lamP) \churchn[2]}
	\breds (\idterm)(\idterm)(\idterm) \fired{(\lamP) \churchn[3]}
	\breds \dots
\end{equation}
where we highlighted the subterms that are actually reduced.
In this latter reduction sequence,
the depth of the reduced redexes actually tends
to infinity, hence by \cref{lem:stratification}
it gives rise to a reduction $\acc \bredi (\idterm)(\idterm)(\idterm)\ldots
= \acc*$. Too bad!

In order to ensure that \emph{no} reduction sequence starting from $\acc$ can
be turned into a valid reduction $\acc \bredi \acc*$,
we will follow a similar pattern, but ensure that:
\begin{itemize}
    \item (easy) \(B_0\) is distinct from the other \(B_i\)'s,
        so that it can act as a marker for the root of
        the comb shaped tree \(\acc*\), and of its approximations
        along the reduction;
    \item (the hard part) no reduction sequence below the root of the tree
        \((B_0)\dots(B_{n})C_{n}\) can build \((B_0)\dots(B_{n+1})C_{n+1}\)
        because, intuitively, the machinery in the lower
        part of the tree needs an information produced by
        the reduction of the root redex involving \(B_0\)
        before it can build \((B_0)\dots(B_{k})C_{k}\)
        with \(k>n\).
\end{itemize}
Our solution is as follows:

\begin{defi} \label{defi:acc}
	The \defemph{Accordion} λ-term is defined as
	$\acc \eqdef (\lamP) \churchn[0]$, where:
	\[
		\lamP \eqdef (\yfp)\, λp. λ n.\, (\appterm 
		\ttterm)\, ((n) \appterm \ffterm)\, \lamQ_{p, n} \qquad
		\lamQ_{p,n} \eqdef (\yfp)\, λq. λ b.\,
			((b) (p) (\churchsucc) n)\, q.
	\]
	We also define
	$\acc* \eqdef (\appterm \ttterm)\infrapp{\appterm \ffterm}$.
\end{defi}

The key in this definition is that it does ensure that
$C_n \eqdef \lamQ_{\lamP, \churchn}$
(or, in practice, some β-equivalent term)
needs to interact with  $B_0 \eqdef \appterm{\ttterm}$,
to be able to produce the next $B_{n+1}$ and $C_{n+1}$.
Therefore there is no other way to produce better and better
approximations of $\acc*$ than the following:
\begin{equation} \label{eq:acc:acc-illustration}
\begin{tikzpicture}[
baseline, 
every node/.style={anchor=base}, 
xscale=0.95, yscale=0.7
]
	\path	(-1.8, 0)	node (a)	{ $\acc$ };
	
	\path	(-1	, 0)	node (a0fl)	{ $\breds$ };
	
	\path	(0	, 0)	node (0a)	{ $@$ }
	+		(-.3,-1)	node (0p)	{ $\lamP''$ }
	+		(.3	,-1)	node (00)	{ $\churchn[0]$ };
	\draw (0a) -- (0p);
	\draw (0a) -- (00);
	
	\path	(1	, 0)	node (01fl)	{ $\breds$ };

	\path	(2	, 0)	node (1a0)	{ $@$ }
	+		(-.4,-1)	node (1t0)	{ $\appterm \ttterm$ }
	+		(.4	,-1)	node (1q)	{ $\lamQ_0$ };
	\draw (1a0) -- (1t0);
	\draw (1a0) -- (1q);
	
	\path	(3	, 0)	node (12fl)	{ $\breds$ };

	\path	(4	,0)	node (2a)	{ $@$ }
	+		(-.3,-1)	node (2p)	{ $\lamP''$ }
	+		(.3	,-1)	node (21)	{ $\churchn[1]$ };
	\draw (2a) -- (2p);
	\draw (2a) -- (21);
	
	\path	(5	, 0)	node (23fl)	{ $\breds$ };
	
	\path	(6	, 0)	node (3a0)	{ $@$ }
	+		(-.4,-1)	node (3t0)	{ $\appterm \ttterm$ }
	++		(.4	,-1)	node (3a1)	{ $@$ }
	+		(-.4,-1)	node (3t1)	{ $\appterm \ffterm$ }
	+		(.4	,-1)	node (3q)	{ $\lamQ_1$ };
	\draw (3a0) -- (3t0);
	\draw (3a0) -- (3a1);
	\draw (3a1) -- (3t1);
	\draw (3a1) -- (3q);
	
	\path	(7,0)	node (34fl)	{ $\breds$ };
	
	\path	(8	,0)	node (4a)	{ $@$ }
	+		(-.3,-1)	node (4p)	{ $\lamP''$ }
	+		(.3	,-1)	node (4n)	{ $\churchn$ };
	\draw (4a) -- (4p);
	\draw (4a) -- (4n);
	
	\path	(9	,0)	node (45fl)	{ $\breds$ };
	
	\path	(10,0)	node (5a0)	{ $@$ }
	+		(-.4,-1)	node (5t0)	{ $\appterm \ttterm$ }
	++		(.4	,-1)	node (5a1)	{ $@$ }
	+		(-.4,-1)	node (5t1)	{ $\appterm \ffterm$ }
	++		(.4	,-1)	node (5a2)	{ $@$ }
	+		(-.4,-1)	node (5t2)	{ $\appterm \ffterm$ }
	++		(.6	,-1.5)	node (5an)	{ $@$ }
	+		(-.4,-1)	node (5tn)	{ $\appterm \ffterm$ }
	+		(.4	,-1)	node (5q)	{ $\lamQ_n$. };
	\draw 			(5a0) -- (5t0);
	\draw 			(5a0) -- (5a1);
	\draw 			(5a1) -- (5t1);
	\draw 			(5a1) -- (5a2);
	\draw 			(5a2) -- (5t2);
	\draw[dotted] 	(5a2) -- (5an);
	\draw 			(5an) -- (5tn);
	\draw 			(5an) -- (5q);
\end{tikzpicture}
\end{equation}
for given terms $\lamP''$ and $\lamQ_n$ respectively β-equivalent to
$\lamP$ and $\lamQ_{\lamP, \churchn}$.
This dynamics ($\acc$ is \enquote{stretched} and 
\enquote{compressed} over and over)
justifies the name \enquote{Accordion}.
More concretely:
\begin{itemize}
    \item when fed with a Church integer argument \(\churchn\), the term 
    \(\lamP''\)
        produces a term mimicking \(\acc*\) up to the \(n\)-th copy of
        \(\appterm\ffterm\),
        the latter being applied to \(\lamQ_n\);
    \item the applicator \(\appterm-\) enforces a kind of call-by-value
        discipline, giving control to the argument (observe that \((\appterm 
        M)N\bred(N)M \));
    \item \(\lamQ_n\) eats up boolean arguments \(\ffterm\), 
        until it is fed with a boolean \(\ttterm\) (marking the root of the 
        tree),
        at which point it restores \(\lamP''\), applied to the next Church 
        integer.
\end{itemize}

\begin{thm} \label{thm:acc}
	\begin{ienumerate}
	\item $\Tay(\acc) \Lrred \Tay(\acc*)$, but
	\item there is no reduction $\acc \bredi \acc*$.
	\end{ienumerate}
\end{thm}

This theorem improves on the results from the first author's
PhD~thesis \cite[Theorem~5.12]{CerdaPhD},
where only the qualitative setting was treated
(\ie when $\SSS = \mathbb{B}$).
Non-conservativity in the general case was presented
as Conjecture~5.15, which is thereby solved.

\subsection{Proof of the counterexample}
\label{sec:acc:acc-proof}

In this (essentially technical) section, we prove \cref{thm:acc}:
a reader already satisfied with the above intuitions might prefer to skip it,
and jump to \cref{sec:uniformity}.
The key ingredients in the proof are
the following well-known notions as well as
the associated factorization property
due to Mitschke~\cite[cor.~5]{Mitschke79}.

\begin{defi} \label{defi:head}
	A λ-term $M \in \lci$ has two possible \defemph{head forms}:
	\begin{itemize}
	\item either the form $λx_1\dots λx_m.(y) M_1 \dots M_n$,
		called \defemph{head normal form} (\textsc{hnf}),
	\item or the form $λx_1\dots λx_m.(λx.P) Q M_1 \dots M_n$,
		where $(λx.P) Q$ is called the \defemph{head redex}.
	\end{itemize}
	As a consequence, a β-reduction $M \bred N$ reduces:
	\begin{itemize}
	\item either a head redex: it is a \defemph{head reduction},
		denoted by $M \hred N$,
	\item or any other redex: it is an \defemph{internal reduction},
		denoted by $M \ired N$.
	\end{itemize}
\end{defi}

\begin{lem}[head-internal decomposition]
\label{lem:head-internal}
	For all $M,N \in \lc$ such that $M \breds N$,
	there exists an $M' \in \lc$ such that
	$M \hreds M' \ireds N$.
\end{lem}

The proofs of both implications of \cref{thm:acc}
rely on the following basic ideas:
\begin{enumerate}[label=(\roman*)]
\item We show that $\acc$ can be (head) reduced to arbitrarily accurate
	approximations of $\acc*$.
	As a consequence, each approximant in $\Tay(\acc*)$
	is in the reduct of an approximant in $\Tay(\acc)$.
	After some technical work, we can conclude that indeed
	$\Tay(\acc) \Lrred \Tay(\acc*)$.
\item If there is a reduction $\acc \bredi \acc*$,
	then by \cref{lem:head-internal}
	there is such a reduction made of head reductions
	followed only by internal reductions.
	By examining all the possible head reducts of $\acc*$,
	we show that it is not possible to produce $\acc*$
	in this way.
\end{enumerate}
For both directions, we first need to write the complete
head reduction sequence starting from $\acc$: let us do this now.
We will use the following abbreviations\footnote{%
	Notice that the $\lamQ_n$ we define here
	are slightly different from those in the example reduction
	from \cref{eq:acc:acc-illustration},
	but they are β-equivalent and play the same role.
}:
\begin{gather*}
	\lamP'		\eqdef λp. λ n.\, (\appterm \ttterm)\,
				((n) \appterm \ffterm)\, \lamQ_{p, n} \qquad
	\lamP''		\eqdef \left( λ x.\,(\lamP')(x)x \right)\, 
				λ x.(\lamP')(x)x \qquad
	\lamQ_n 	\eqdef \lamQ_{\lamP'',(\churchsucc)^n \churchn[0]} \\
	\lamQ'_n	\eqdef λq. λ b.\, ((b) (\lamP'') 
				(\churchsucc)^{n+1}\churchn[0])	\, q \qquad
	\lamQ''_n	\eqdef (λ x.(\lamQ'_n)(x)x) \, λx.(\lamQ'_n)(x)x.
\end{gather*}
The first step is:
\newcommand{\hRED}{\makebox[3em]{\ensuremath{\hred}}}
\newcommand{\hREDS}{\makebox[3em]{\ensuremath{\hreds}}}
	\begin{equation} \label{eq:acc-hred:debut}
		\acc = ( \fired{ (\yfp) \lamP' } ) \churchn[0]
		\hred (\lamP'') \churchn[0]
	\end{equation}
	Then, for each $n \in \Nats$, we do the following head reduction steps:
	\begin{align}
		  & (\fired{\lamP''}) (\churchsucc)^n \churchn[0] \notag \\
			\label{eq:acc-hred:boucle}
	\hRED & \left( \fired{ (\lamP') \lamP'' } \right) (\churchsucc)^n 
			\churchn[0]	\\
	\hRED &	\fired{ \left( λ n.\, (\appterm\ttterm)\, ((n) 
			\appterm\ffterm)\, 
			\lamQ_{\lamP'', n} \right)\, (\churchsucc)^n \churchn[0] } 
			\label{eq:acc-hred:redex1}	\\
	\hRED &	\fired{ (\appterm\ttterm)\, \left( \left( (\churchsucc)^n 
			\churchn[0] \right)	\appterm\ffterm \right)\, \lamQ_n } 
			\label{eq:acc-hred:redex2}	\\
	\hRED & \fired{ (\churchsucc)^n \churchn[0] } \; \appterm\ffterm \, 
			\lamQ_n \, \ttterm	\\
	\hRED &	\fired{ \left( λ f.λ x. (( (\churchsucc)^{n-1}
			\churchn[0] )f)(f)x \right) \appterm\ffterm }
			 \, \lamQ_n \, \ttterm
			\label{eq:acc-hred:succdeb} \\
	\hRED &	\fired{ \left( λ x. (( (\churchsucc)^{n-1} \churchn[0] )
			\appterm\ffterm) (\appterm\ffterm) x \right) \lamQ_n } \, 
			\ttterm \\
	\hRED &	\left( \fired{ (\churchsucc)^{n-1} \churchn[0] } \, 
			\appterm\ffterm\, (\appterm\ffterm)\lamQ_n \right) \ttterm	
			\label{eq:acc-hred:succfin} \\
	\intertext{and by repeating steps (\ref{eq:acc-hred:succdeb}) to 
	(\ref{eq:acc-hred:succfin}):}
	\hREDS& \left( \fired{(\churchn[0]) \appterm\ffterm} \, 
			(\appterm\ffterm)^n\lamQ_n \right) \ttterm \\
	\hRED & \left( \fired{ (λ x.x) \, (\appterm\ffterm)^n\lamQ_n } \right) 
			\ttterm \\
	\hRED &	\left( \fired{ (λ b.(b)\ffterm) \, 
				(\appterm\ffterm)^{n-1} \lamQ_n }
			\right) \ttterm	\\
	\hRED & \left( \fired{ (\appterm\ffterm)^{n-1} \lamQ_n } \right) \ffterm 
			\, \ttterm \label{eq:acc-hred:lff}
	\end{align}
	and by repeating step (\ref{eq:acc-hred:lff}):
	\begin{align}
	\hREDS& \left( \fired{(\yfp) \lamQ'_n} \right)
			\underbrace{\ffterm \dots \ffterm}_{\text{$n$ times}} \ttterm \\
	\hRED &	( \fired{\lamQ''_n} ) \, \ffterm \dots \ffterm \, \ttterm \\
	\hRED &	\left( \fired{ (\lamQ'_n) \lamQ''_n } \right) \ffterm \dots 
			\ffterm \, \ttterm \label{eq:acc-hred:qdeb} \\
	\hRED &	\fired{ \left( λ b.\, ((b) (\lamP'') 
			(\churchsucc)^{n+1}\churchn[0]) \, 
			\lamQ''_n \right) \ffterm } \dots \ffterm \, \ttterm \\
	\hRED & \left( \left( \fired{
			(λ x.λ y.y) (\lamP'') (\churchsucc)^{n+1}\churchn[0] 
			} \right) \, \lamQ''_n \right)
			\underbrace{\ffterm\dots\ffterm}_{\substack{n-1\\\text{times}}} 
			\ttterm \\
	\hRED &	\left( \fired{(λ y.y) \lamQ''_n} \right) \ffterm \dots \ffterm\, 
			\ttterm \\
	\hRED &	( \fired{\lamQ''_n} ) \, \ffterm \dots \ffterm \, \ttterm 
			\label{eq:acc-hred:qfin}	\\
	\intertext{and by repeating steps (\ref{eq:acc-hred:qdeb}) to 
	(\ref{eq:acc-hred:qfin}):}
	\hREDS& ( \fired{\lamQ''_n} ) \, \ttterm \\
	\hRED &	\left( \fired{ (\lamQ'_n) \lamQ''_n } \right) \ttterm	\\
	\hRED &	\fired{ \left( λ b.\, ((b) (\lamP'') 
			(\churchsucc)^{n+1}\churchn[0]) \, \lamQ''_n \right) \ttterm } 
			\label{eq:acc-hred:redex3} \\
	\hRED & \left( \fired{(λ x.λ y.x) (\lamP'') 
			(\churchsucc)^{n+1}\churchn[0]} \right) \, \lamQ''_n \\
	\hRED &	\fired{ \left( λ y. (\lamP'') \, (\churchsucc)^{n+1}\churchn[0] 
			\right) \lamQ''_n } \label{eq:acc-hred:redex4} \\
	\hRED &	(\lamP'') \, (\churchsucc)^{n+1}\churchn[0]
		 \label{eq:acc-hred:fin}
	\end{align}
	which brings us back to step (\ref{eq:acc-hred:boucle}).

\medskip

We are now able to start the proof of \cref{thm:acc},
whose two parts are the content of \cref{lemma:acc-i,lemma:acc-ii}.
Notice that in the proof of \cref{lemma:acc-i} below,
one technical argument relies on material to be introduced
in \cref{sec:uniformity:definitions}:
the reader reluctant to put a temporary faith in our assertions
may want to have a look there first.

\begin{lem}[\cref{thm:acc}, item (i)] \label{lemma:acc-i}
	There is a reduction $\Tay(\acc) \Lrred \Tay(\acc*)$.
\end{lem}

\begin{proof}
 	For all $d \in \Nats$, we define:
 	$\acc*_d \eqdef (\appterm{\ttterm}) (\appterm{\ffterm})^d \lamQ_n$.
	As a consequence of the reduction described in 
	\crefrange{eq:acc-hred:debut}{eq:acc-hred:fin},
	in particular its step \ref{eq:acc-hred:redex2},
	there are reductions
	$\acc \breds \acc*_0 \breds \acc*_1 \breds \acc*_2 \breds \dots$
	By \cref{thm:Lrred-simul-breds}, we obtain
	\begin{equation} \label{eq:acc:proof-it1:1}
		\Tay(\acc) \Lrred \Tay(\acc*_0) \Lrred \Tay(\acc*_1) 
		\Lrred \Tay(\acc*_2) \Lrred \dots
	\end{equation}
 	For all $d \in \Nats$, we also define
 	$\Tay_{d}(\acc*) \suminclusion \Tay(\acc*)$
 	to be the \enquote{sub-sum}
 	containing only the approximants of applicative depth~$d+1$.
 	Explicitely, we first define $\Tay'_d(\acc*) \eqdef 
	\Tay( (\appterm{\ttterm}) (\appterm{\ffterm})^d \bot )$,
	where $\bot$ is a constant such that $\Tay(\bot) \eqdef 0$
	(this is just a trick to \enquote{cut} the Taylor expansion
	at some point), and then 
	\[	\left\{ \begin{aligned}
		\Tay_0(\acc*) & \eqdef \Tay'_0(\acc*) \\
		\Tay_{d+1}(\acc*) & \eqdef \Tay'_{d+1}(\acc*) - \Tay'_d(\acc*)
			= \sum\nolimits_{ s \in \sumsupp{\Tay'_{d+1}(\acc*)}
				\setminus \sumsupp{\Tay'_d(\acc*)}
			} \Tay(s,\acc*) \cdot s.
		\end{aligned} \right. \]
	
	Let us make the following two sec:preliminary observations.
	\begin{itemize}
	\item By construction (using the observation that the coefficient
		of $s \in \sumsupp{\Tay(M)}$ does not depend on $M$),
		we obtain:
		\begin{gather}
			\Tay(\acc*_d) = \Tay_{d}(\acc*) + \SS_d,
				\text{ for some $\SS_d$ such that }
				\sumsupp{\Tay_d(\acc*)} \cap \sumsupp{\SS_d} = \varnothing
			\label{eq:acc:proof-it1:2} \\
			\Tay(\acc*) = \sum_{n \in \Nats} \Tay_{d}(\acc*)
				\label{eq:acc:proof-it1:3}
		\end{gather}
	\item In addition,
		\begin{equation} \label{eq:acc:proof-it1:4}
			\forall s \in \Tay_d(\acc*),\ \forall k > 0,\ 
			\nexists t \in \Tay_{d+k}(\acc*),\ 
			s \rreds t + T
		\end{equation}
		for some $T \in \frsums$:
		this is due to the fact that terms in $\Tay(\acc*)$
		cannot see their (applicative) depth increase 
		through resource reduction.
		In particular it means that in any reduction
		\[	\Tay(\acc*_d) = \Tay_{d}(\acc*) + \SS_d \Lrred 
			\Tay(\acc*_{d+k}) = \Tay_{d+k}(\acc*) + \SS_{d+k}, \]
		only the terms of $\SS_d$ actually contribute to $\Tay_{d+k}(\acc*)$.
		This is the case in particular for the reductions
		coming from \cref{eq:acc:proof-it1:1}, as we will now consider.
	\end{itemize}
	
	Now all the material and hypotheses have been exposed,
	the proof goes as follows.
	We want to define a sequence of sums
	$\SS'_d \suminclusion \SS_d \suminclusion \Tay(\acc*_d)$
	such that for all $d \in \Nats$,
	$\SS'_d \Lrred \Tay_{d+1}(\acc) + \SS'_{d+1}$
	and such that in all reductions
	$\Tay(\acc*_d) \Lrred \Tay(\acc*_{d+k})
	= \Tay_{d+k}(\acc*) + \SS_{d+k}$
	coming from \cref{eq:acc:proof-it1:1,eq:acc:proof-it1:2}, for $k > 0$,
	only the terms of $\SS'_d$ contribute to $\Tay_{d+k}(\acc)$.
	\begin{itemize}
	\item For $d = 0$, take $\SS'_0 \eqdef \SS_0$.
		By \cref{eq:acc:proof-it1:2}, $\Tay(\acc*_0) = \Tay_0(\acc*) + \SS_0$
		and the desired property is a direct consequence of
		\cref{eq:acc:proof-it1:4}.
	\item For $d \geq 0$, consider the reduction
		$\Tay(\acc*_d) \Lrred \Tay(\acc*_{d+1})
		= \Tay_{d+1}(\acc*) + \SS_{d+1}$
		coming from \cref{eq:acc:proof-it1:1,eq:acc:proof-it1:2}.
		Suppose that $\SS'_d$ is built, then because we know that
		only the terms from $\SS'_d$ contribute to $\Tay_{d+1}(\acc*)$
		in the above reduction we can decompose it as follows:
		\begin{equation} \label{eq:acc:proof-it1:5}
			\Tay(\acc*_d) = \left\{ \begin{array}{c@{}r@{}}
				\TT_1 & \Lrred 
				\\ + \\
				\SS'_d & \mbox{} = \left\{ \begin{array}{cr@{}}
					\TT_2 & \Lrred 
					\\ + \\
					\TT_3 & \Lrred 
				\end{array} \right.
			\end{array} \right.
			\left. \begin{array}{@{}c@{}c}
				\left. \begin{array}{c}
					\TT_4
					\\ + \\
					\SS'_{d+1}
				\end{array} \right\} = & \SS_{d+1}
				\\& + \\&
				\Tay_{d+1}(\acc*)
			\end{array} \right\} = \Tay(\acc*_{d+1})
		\end{equation}
		for some $\TT_1, \dots, \TT_4 \in \irsums$
		such that $\SS'_{d+1}$ is defined to be the reduct of $\TT_2$.
		In addition, for all $k > 0$:
		\begin{itemize}
		\item in the reduction
			$\Tay(\acc*_{d+1}) \Lrred \Tay(\acc*_{d+1+k}) 
			= \Tay_{d+1+k}(\acc*) + \SS_{d+1+k}$
			coming from \cref{eq:acc:proof-it1:1,eq:acc:proof-it1:2},
			by the observation (\ref{eq:acc:proof-it1:4}),
			only the terms from $\SS_{d+1}$ contribute
			to $\Tay_{d+1+k}(\acc*)$,
		\item in the reduction
			$\Tay(\acc*_{d}) \Lrred \Tay(\acc*_{d+1+k}) 
			= \Tay_{d+1+k}(\acc*) + \SS_{d+1+k}$
			also coming from \cref{eq:acc:proof-it1:1,eq:acc:proof-it1:2},
			by the property ensured on $\SS'_d$,
			only the terms from $\SS'_d$ contribute
			to $\Tay_{d+1+k}(\acc*)$,
		\end{itemize}
		and the latter reduction is obtained by appending the former
		to the reduction (\ref{eq:acc:proof-it1:5}).
		As $\SS'_{d+1}$ contains exactly the terms of $\SS_{d+1}$
		coming from $\SS'_d$,
		only the terms of $\SS'_{d+1}$ contribute to
		$\Tay_{d+1+k}(\acc*)$ in the former reduction,
		which was the desired property.
	\end{itemize}
	In the end, we obtain, for all $N \in \Nats$:
	\begin{equation} \label{eq:acc:proof-it1:seqLrred}
		\letrelsbreathe
		\Tay(\acc) \Lrred \Tay_0(\acc*) + \SS_0
		\Lrred \Tay_0(\acc*) + \Tay_1(\acc*) + \SS'_1
		\Lrred \dots \Lrred \sum_{d=0}^{N} \Tay_d(\acc*) + \SS'_N
	\end{equation}
	For each $s \in \sumsupp{\Tay(\acc)}$,
	this can be turned into:
	\begin{equation} \label{eq:acc:proof-it1:seqrreds}
		\letrelsbreathe
		s \rreds T_{s,0} + S_{s,0} \rreds T_{s,0} + T_{s,1} + S_{s,1}
		\rreds \dots \rreds \sum_{d=0}^{N} T_{s,d} + S_{s,N}
	\end{equation}
	for some $T_{s,d}, S_{s,d} \in \frsums$ satisfying
	$\Tay_d(\acc*) = \sum_{s \in \rc} \Tay(s, \acc) \cdot T_{s,d}$.
	At this point the careful reader might raise an eyebrow,
	because taking a lifted reduction $\Lrred$ in $\irsums$
	back to reductions $\rreds$ in $\frsums$
	may not be possible in general.
	The reason for this is that 
	the latter only works with integer coefficients
	whereas we could start from a reduction like
	$s \Lrred \frac 13 \cdot s + \frac 23 \cdot s$ in $\irsums[\Rationals]$.
	This is where we need the material to be introduced
	in \cref{sec:uniformity:definitions}:
	in fact the reductions $\Lrred$ from \cref{eq:acc:proof-it1:1}
	were obtained \emph{via} the simulation \cref{thm:Lrred-simul-breds},
	but the refined \emph{uniform} simulation \cref{thm:cohrred-simul-bred}
	allows to replace these reductions with $\cohrreds$.
	As a consequence, all reductions $\Lrred$
	from \cref{eq:acc:proof-it1:1} to \cref{eq:acc:proof-it1:seqLrred}
	are in fact uniform reductions $\cohrreds$.
	In addition for all $d \in \Nats$
	the sums $\Tay_d(\acc*)$ and $\SS'_d$ have disjoint supports,
	hence \cref{eq:acc:proof-it1:seqrreds}
	is a consequence of \cref{lem:cohrred-from-sums-to-fsums}
	\footnote{Notice that obtaining \cref{eq:acc:proof-it1:seqrreds}
		from \cref{eq:acc:proof-it1:seqLrred} is immediate
		when $\SSS$ is the semiring of booleans,
		as was done in \cite{CerdaPhD},
		and remains possible if $\SSS$ has the \enquote{refinement}
		or \enquote{additive splitting} property,
		which is the case of all semirings used in practice
		to our knowledge;
		we rely on \cref{lem:cohrred-from-sums-to-fsums}
		only to provide the most general proof possible.
	}.
	
	We are now ready to conclude. Observe the following facts:
	\begin{itemize}
	\item For each $s \in \Tay(\acc)$
		there are only finitely many $d \in \Nats$
		such that $T_{s,d} \neq 0$.
		This is due to the fact that a resource term
		has only finitely many reducts \cite[Lemma~3.13]{Vaux19}.
	\item $\acc$ has no head normal form
		as demonstrated in 
		\crefrange{eq:acc-hred:debut}{eq:acc-hred:fin},
		which entails that $\Tay(\acc) \Lrred 0$
		\cite[Theorem~5.6]{CerdaVauxAuclair23}.
		Since $\SS'_N$ only contains reducts of terms in $\Tay(\acc)$,
		this means that we can reduce $\SS'_N \rreds 0$.
	\end{itemize} 
	As a consequence, $s \rreds \sum_{d \in \Nats} T_{s,d}$
	and therefore:
	\[	\letrelsbreathe
		\Tay(\acc) = \sum_{s \in \rc} \Tay(s,\acc) \cdot s
		\Lrred \sum_{s \in \rc} \Tay(s,\acc) \cdot 
			\sum_{d \in \Nats} T_{s,d}
		= \sum_{d \in \Nats} \Tay_d(\acc*)
		= \Tay(\acc*) \]
	by \cref{eq:acc:proof-it1:3}.
\end{proof}

We can now start the second part of the proof of the counterexample.
It consists in showing that the undesired behaviour
illustrated on our naive counterexample candidate
(see \cref{eq:acc:bad-counterexample})
cannot be reproduced with $\acc$:
all reduction paths have an \enquote{accordion-like} behaviour,
hence do not correspond to an infinitary β-reduction.

\begin{lem}[\cref{thm:acc}, item (ii)] \label{lemma:acc-ii}
	There is no reduction $\acc \bredi \acc*$.
\end{lem}

 \begin{proof}
	We suppose that there is a reduction
	$\acc \bredi \acc*$ and we show that this leads
	to a contradiction. 
	By \cref{lem:stratification,lem:head-internal},
	there exists respectively a sequence of terms $\acc_d \in \lc$
	and a term $\acc'_0 \in \lc$
	such that there are reductions
	\[
		\acc \hreds \acc'_0 \ireds \acc_1 \breds[≥1] \acc_d
		\bredi[≥d] \acc*.
	\]
	$\acc'_0$ and $\acc*$ must have the same head form,
	\ie there must be $M,N \in \lc$ such that $\acc'_0 = (λb.M)N$.
	The exhaustive description of the head reducts of $\acc$
	detailed in 
	\crefrange{eq:acc-hred:debut}{eq:acc-hred:fin} allows to 
	observe that 
	this only happens in four cases
	(corresponding to steps \ref{eq:acc-hred:redex1},
	\ref{eq:acc-hred:redex2},
	\ref{eq:acc-hred:redex3} and
	\ref{eq:acc-hred:redex4} in 
	\crefrange{eq:acc-hred:debut}{eq:acc-hred:fin}):
	\begin{enumerate}
	\item $\acc'_0 = 
		\left( λ n.\, (\appterm\ttterm)\, ((n) \appterm\ffterm)\, 
		\lamQ_{\lamP'', n} \right)\, (\churchsucc)^n \churchn[0]$,
	\item $\acc'_0 = 
		(\appterm\ttterm)\, \left( \left( (\churchsucc)^n \churchn[0] 
		\right) \appterm\ffterm \right)\, \lamQ_n$,
	\item $\acc'_0 = 
		\left( λ b.\, ((b) (\lamP'') (\churchsucc)^{n+1}\churchn[0]) \, 
		\lamQ''_n \right) \ttterm$,
	\item $\acc'_0 = 
		\left( λ y. (\lamP'') \, (\churchsucc)^{n+1}\churchn[0] 
		\right) \lamQ''_n$,
	\end{enumerate}
	for some $n \in \Nats$ (in the following, $n$ denotes this
	specific integer appearing in $\acc'_0$). In particular,
	for one of these possible values of $\acc'_0$
	there must be a reduction
	\[ \acc'_0 \ireds \acc_{n+4} \bredi[≥n+4] \acc*. \]
	Since $\acc_{n+4}$ and $\acc*$ are identical 
	up to applicative depth $n+3$,
	we can write
	$\acc_{n+4} = (\appterm\ttterm)(\appterm\ffterm)^{n+1} M$
	for some $M \in \lc$ such that
	$M \bredi \infrapp{\appterm\ffterm}$
	(we need to go up to depth $n+3$
	since $\appterm\ttterm$ and $\appterm\ffterm$ are themselves of 
	applicative depth~2).
	Finally, there must be a reduction
	\[ \acc'_0 \ireds (\appterm\ttterm)(\appterm\ffterm)^{n+1} M. \]
	For each of the possible cases for $\acc'_0$,
	let us show
	that this is impossible. The easy cases are:
	\begin{description}
	\item[Case 1, step (\ref{eq:acc-hred:redex1})] 
		Such a reduction would imply that
		$(\churchsucc)^n \churchn[0] \breds (\appterm\ffterm)^{n+1} M$.
		However $(\churchsucc)^n \churchn[0] \breds \churchn$,
		which is in β-normal form,
		while $(\appterm\ffterm)^{n+1} M$ has no normal form.
		We conclude by confluence of the finite λ-calculus.
	\item[Case 3, step (\ref{eq:acc-hred:redex3})]
		Immediate because $\ttterm$ is in normal form.
	\item[Case 4, step (\ref{eq:acc-hred:redex4})]
		Such a reduction would imply that 
		$	λy.(\lamP'')(\churchsucc)^{n+1}\churchn[0]
			\breds \appterm{\ttterm} = λy.(y)\ttterm, $
		and therefore that $(\lamP'')(\churchsucc)^{n+1}\churchn[0]$
		has a \textsc{hnf} $(y)\ttterm$.
		This is impossible, as detailed in the exhaustive
		head reduction of $\acc$ in 
		\crefrange{eq:acc-hred:debut}{eq:acc-hred:fin}.
	\end{description}
	
	The remaining case concerns the reduct 
	$(\appterm\ttterm)\, \left( \left( (\churchsucc)^n 
	\churchn[0] \right) \appterm\ffterm \right)\, \lamQ_n$.
	It is the only \enquote{non-degenerate} one,
	in the sense that it is where the accordion-like behaviour
	of $\acc$ is actually happening:
	the sub-term $\appterm\ttterm$ here is really \enquote{the same}
	as the one appearing at the root of $\acc*$
	but we need to reduce this sub-term at some point
	(\ie to \enquote{compress} the Accordion).
	Thus there can be no 001-infinitary reduction towards~$\acc*$.
	The formal proof of this case, \ie of the impossibility of
	$(\churchsucc)^n \churchn[0] \appterm\ffterm \lamQ_n \breds 
	(\appterm\ffterm)^{n+1} M$,
	is given by \cref{lem:acc-technicalstuff2} below.
\end{proof}

\begin{lem}[base case of \cref{lem:acc-technicalstuff2}] 
\label{lem:acc-technicalstuff1}
	For all $k\in\Nats$, $n\in\Nats$ and $M \in \lc$, there is no reduction
	\[(\appterm \ffterm)^k\, \lamQ_n \breds (\appterm \ffterm)^{k+1} M.\]
\end{lem}

	\begin{proof}
	We proceed by induction on $k$.
	First, take $k=0$ and suppose there is a reduction $\lamQ_n \breds 
	(\appterm\ffterm) M$.
	By \cref{lem:head-internal}, there are $R,R' \in \lc$
	such that \[\lamQ_n \hreds (λ b.R)R' \ireds (\appterm\ffterm) M
	= (λ b.(b)\ffterm) M.\]
	An exhaustive head reduction of $\lamQ_n$
	gives the possible values of $R$ and $R'$:
	\begin{align*}
		\lamQ_n & \widerel{\hred}{=} (\yfp) \lamQ'_n \\
		& \hred \left( λ x.(\lamQ'_n)(x)x \right) \, λx.(\lamQ'_n)(x)x \\
		& \hred \left( λq. λ b.\, ((b) (\lamP'') 
			(\churchsucc)^{n+1}\churchn[0])	\, q	\right) \lamQ''_n \\
		& \hred λ b.\, ((b) (\lamP'') (\churchsucc)^{n+1}\churchn[0])
			\, \lamQ''_n,
	\end{align*}
	the last reduct being in \textsc{hnf},
    which leaves only the first three possibilities.
	In any of those three cases, $R \breds (b)\ffterm$
	(modulo renaming of $b$ by α-conversion)
	is impossible by immediate arguments,
	so that $(λ b.R)R' \ireds (\appterm\ffterm) M$ cannot hold.
	
	\bigskip
	
	If $k \geq 1$, let us again suppose that there is a reduction
	$(\appterm \ffterm)^k\, \lamQ_n \breds (\appterm \ffterm)^{k+1} M$.
	\cref{lem:head-internal} states that there are $R,R' \in \lc$ 
	such that
	\[(\appterm \ffterm)^k\, \lamQ_n \hreds (λ b.R)R' \ireds 
	(λ b.(b)\ffterm) (\appterm \ffterm)^k M.\]
	An exhaustive head reduction of $(\appterm \ffterm)^k\, \lamQ_n$
	gives the possible values of $R$ and $R'$
	(we write only the reduction steps corresponding to the well-formed 
	reducts --- see the details in the detailed head reduction of $\acc$, 
	steps (\ref{eq:acc-hred:lff}) and following):
	\begin{align*}
		(\appterm \ffterm)^k\, \lamQ_n
		& \widerel{\hreds}{=}
			(λ b.(b)\ffterm) \, (\appterm \ffterm)^{k-1}\, \lamQ_n \\
		& \hreds \left( λ b. \left( (b) (\lamP'') (\churchsucc)^{n+1} 
			\churchn[0] \right) \lamQ''_n \right) \ffterm \\
		& \hreds (λ y.y) \lamQ''_n 	\\
		& \hred \lamQ''_n
	\end{align*}
	In the first case, a reduction 
	$(λ b.(b)\ffterm) \, (\appterm \ffterm)^{k-1}\, \lamQ_n \ireds
	(λ b.(b)\ffterm) (\appterm \ffterm)^k M$
	is impossible because it would imply that
	$(\appterm \ffterm)^{k-1}\, \lamQ_n \breds (\appterm \ffterm)^k M$,
	which is impossible by induction.
	The second and third cases are impossible by immediate arguments;
	the fourth case has already been explored
	($\lamQ''_n$ is exactly the term from the second line
	of the reduction of $\lamQ_n$ above).
	\end{proof}

\begin{lem}[the difficult case of \cref{lemma:acc-ii}] 
\label{lem:acc-technicalstuff2}
	For all $n\in\Nats$, $k \in [0,n]$ and $M \in \lc$,
	there is no reduction:
	\[(\churchsucc)^{n-k}\, \churchn[0]\, \appterm \ffterm\, (\appterm 
	\ffterm)^k \lamQ_n
	\breds (\appterm \ffterm)^{n+1} M. \]
\end{lem}

	\begin{proof}
	We proceed by induction on $n-k$.
	The base case is $k=n$: if there is a reduction
	$(\churchn[0])  \, \appterm \ffterm\, (\appterm \ffterm)^n \lamQ_n
	\breds (\appterm \ffterm)^{n+1} M$,
	then by \cref{lem:head-internal} there are terms $R,R' \in \lc$ 
	such that
	\[(\churchn[0]) \, \appterm \ffterm\, (\appterm \ffterm)^n \lamQ_n
	\hreds (λ b.R)R' \ireds (λ b.(b)\ffterm) (\appterm 
	\ffterm)^n M.\]
	Observe that
	\[
		(\churchn[0]) \, \appterm \ffterm\, (\appterm \ffterm)^n \lamQ_n
		\hred (λ x.x) \, (\appterm \ffterm)^n \lamQ_n
		\hred (\appterm \ffterm)^n \lamQ_n
	\]
	hence, because $λ x.x$ is in β-normal form
	and by \cref{lem:acc-technicalstuff1}, we reach a contradiction.
	
	\bigskip
	
	If $k < n$ and there is a reduction 
	$(\churchsucc)^{n-k}\, \churchn[0]\, \appterm \ffterm\, (\appterm 
	\ffterm)^k \lamQ_n
	\breds (\appterm \ffterm)^{n+1} M$, then again by 
	\cref{lem:head-internal}
	there are terms $R,R' \in \lc$ such that
	\[(\churchsucc)^{n-k}\, \churchn[0]\, \appterm \ffterm\, (\appterm 
	\ffterm)^k \lamQ_n
	\hreds (λ b.R)R' \ireds (λ b.(b)\ffterm) (\appterm 
	\ffterm)^n M.\]
	Observe that
	\begin{align*}
		(\churchsucc)^{n-k}\, \churchn[0]\, \appterm \ffterm\, (\appterm 
		\ffterm)^k \lamQ_n
		& \hred \left( λ f.λ x.(\churchsucc)^{n-k-1} \, 
		\churchn[0] \,
			f \, (f) x \right) \, \appterm\ffterm \, (\appterm\ffterm)^k 
			\lamQ_n			
			\\
		& \hred \left( λ x.(\churchsucc)^{n-k-1} \, \churchn[0] \, 
		\appterm\ffterm \, 
			(\appterm\ffterm) x \right) \, (\appterm\ffterm)^k 
			\lamQ_n						
			\\
		& \hred (\churchsucc)^{n-k-1}\, \churchn[0]\, \appterm \ffterm\,
			(\appterm \ffterm)^{k+1} \lamQ_n
	\end{align*}
	The first reduct does not have the expected head form.
	In the second case,
	$(λ b.R)R' \ireds (λ b.(b)\ffterm) (\appterm \ffterm)^n M$
	would imply that $(\appterm\ffterm)^k \lamQ_n \breds (\appterm 
	\ffterm)^n M$,
	which is impossible by \cref{lem:acc-technicalstuff1} because $k < n$.
	In the third case, apply the induction hypothesis.
	\end{proof}

%*******************************************************************************

\section{The missing ingredient: Uniformity} \label{sec:uniformity}

The fact that the simulation of $\bredi$ by $\Lrred$
\emph{via} Taylor expansion is not conservative
confirms that pointwise reduction $\Lrred$,
even if needed in order to express the pointwise normal form
of a sum through resource reduction,
weakens the dynamics of the β-reduction
by allowing to reduce resource approximants
along reductions paths that do not correspond
to an actual reduction of the approximated term.
As already underlined by Ehrhard and Regnier in their
seminal work \cite{EhrhardRegnier08},
\emph{uniformity} is what gives the linear approximation
all its robustness; this will also be the case for our study.

We first recall from \cite{EhrhardRegnier08,CerdaPhD}
the definition of a coherence relation on resource terms
such that Taylor expansions of λ$\bot$-terms are uniform
(\ie self-coherent),
and the fact that resource reduction can be uniformly lifted
to such uniform infinite sums,
yielding a uniform simulation of the infinitary β-reduction
(\cref{sec:uniformity:definitions}).
Then we prove that this simulation,
contrary to the simulation by the weaker lifting $\Lrred$,
is conservative (\cref{sec:uniformity:conservativity}).

\subsection{Uniform simulation of the infinitary β-reduction}
\label{sec:uniformity:definitions}

The intuition behind coherence and uniformity is simple:
two resource terms are coherent with each other when their syntax trees have the same shape,
and in particular, in argument positions, all elements of both
argument multisets are pairwise coherent, for instance:
\[	x \coh x \not\coh y \qquad 
	x[λx.x[y], λx.x1] \coh x[λx.x[y,y,y]] ; \]
this extends to sums by saying that two sums are coherent
whenever all pairs of elements of their supports are coherent;
a resource term (or sum) is uniform when it is self-coherent.
Formally:

\begin{defi} \label{defi:coh}
	The relation
	$\mathord{\coh} \subset \rc** \times \rc**$
	of \defemph{coherence} is defined by the rules:
	\begin{prooftreeset}
		\begin{prooftree}
			\infer0[\Vars_{\coh}]{ x \coh x }
		\end{prooftree}
	\qquad
		\begin{prooftree}
			\hypo{ s \coh s' }
			\infer1[λ_{\coh}]{ λx.s \coh λx.s' }
		\end{prooftree}
	\qquad
		\begin{prooftree}
			\hypo{ s \coh s' }
			\hypo{ \ms t \coh \ms t' }
			\infer2[@_{\coh}]{ \rapp st \coh \rapp st' }
		\end{prooftree}
	\\[\topsep]
		\begin{prooftree}
		\hypo{ \forall i \in \{1,\dots,m\},\ 
			\forall j \in \{1,\dots,n\},\ t_i \coh t'_j }
		\infer1[!_{\coh}]{ [t_1,\dots,t_m] \coh [t'_1,\dots,t'_n] }
		\end{prooftree}
	\end{prooftreeset}
	For $\SS,\TT \in \irsums**$, we write $\SS \coh \TT$
	whenever $\forall s \in \sumsupp{\SS},\ 
	\forall t \in \sumsupp{\TT},\ s \coh t$.
\end{defi}

We call \defemph{uniform} any $\SS \in \irsums$ such that $\SS \coh \SS$.
A crucial observation is that for all $M \in \lbci$,
$\Tay(M)$ is uniform by construction:
all its elements have \enquote{the shape of (a prefix of) $M$}.

We now introduce a uniform lifting $\cohrred$
of the resource reduction $\rred$. Intuitively:
\begin{itemize}
\item this lifting can only reduce uniform sums, 
\item each uniform reduction step of a sum
	is a \enquote{bundle} of resource reduction steps
	occurring at the same address in the elements of the sum
	(\(\cohOrred\) is an inductive reformulation
	of Midez' \enquote{giant-step} \(Γ\)-reduction,
	whose definition uses explicit addresses~\cite{Midez14}).
\end{itemize}
This allows to capture only the reductions of some $\Tay(M)$
that correspond to a β-reduction of $M$,
as will be expressed by the conservative simulation stated
in \cref{thm:cohrred-simul-bred,thm:conserv-cohrredi-bredi} below.

\begin{defi} \label{defi:cohrred}
	Given an index set $I$, we define an auxiliary relation
	$\mathord{\cohOrred} \subset (\rc**)^I \times (\frsums**)^I$
	by the following rules:
	\begin{prooftreeset}
		\begin{prooftree}
		\hypo{ \forall i,j,\ s_i \coh s_j  }
		\hypo{ \forall i,j,\ \ms t_i \coh \ms t_j }
		\infer2[beta_{\cohresource}]
			{ (\rapp{λx.s_i}t_i)_{i\in I} 
			\cohOrred (\rsubst*{s_i}{\ms t_i})_{i\in I} }
		\end{prooftree}
	\qquad
		\begin{prooftree}
		\hypo{ (s_i)_{i\in I} \cohOrred (S'_i)_{i\in I} }
		\infer1[λ_{\cohresource}]
			{ (λx.s_i)_{i\in I} \cohOrred (λx.S'_i)_{i\in I} }
		\end{prooftree}
	\\[\topsep]
		\begin{prooftree}
		\hypo{ (s_i)_{i\in I} \cohOrred (S'_i)_{i\in I} }
		\hypo{ \forall i,j,\ \ms t_i \coh \ms t_j }
		\infer2[@l_{\cohresource}]{ (\rapp{s_i}t_i)_{i\in I} 
			\cohOrred (\rapp{\smash{S'_i}}t_i)_{i\in I} }
		\end{prooftree}
	\qquad
		\begin{prooftree}
		\hypo{ \forall i,j,\ s_i \coh s_j  }
		\hypo{ (\ms t_i)_{i\in I} \cohOrred (\ms T'_i)_{i\in I} }
		\infer2[@r_{\cohresource}]{ (\rapp{s_i}t_i)_{i\in I} 
			\cohOrred (\rapp{s_i}T'_i)_{i\in I} }
		\end{prooftree}
	\\[\topsep]
		\begin{prooftree}
			\hypo{ (t_{i,j})_{\begin{subarray}{l}
					i\in I \\ 1\le j\le k_i
				\end{subarray}} \cohOrred (T'_{i,j})_{\begin{subarray}{l}
					i\in I \\ 1\le j\le k_i
				\end{subarray}} }
			\infer1[!_{\cohresource}]
				{ ([ t_{i,1},\dotsc, t_{i,k_i} ])_{i\in I} \cohOrred 
				([ T'_{i,1},\dotsc, T'_{i,k_i} ])_{i\in I} }
		\end{prooftree}
	\end{prooftreeset}
	and the relation
	$\mathord{\cohrred} \subset \irsums** \times \irsums**$
	of \defemph{uniform resource reduction}
	is defined by:
	\[
		\begin{prooftree}
		\hypo{ (u_i)_{i\in I} \cohOrred (U'_i)_{i\in I} }
		\infer1[\Sigma_{\cohresource}]{ \sum_{i \in I} a_i u_i \cohrred 
			\sum_{i \in I} a_i U'_i. }
		\end{prooftree}
	\]
	
	As we are now acquainted with, we also define the relation
	$\mathord{\cohrred[≥d]} \subset \irsums** \times \irsums**$
	of \defemph{uniform resource reduction at minimum depth \textit{d}},
	for $d \in \Nats$,
	by the following indexed version of the above rules:
	
	\vspace{\prooftreevsep}\vspace{\topsep}
	\begin{widecenter}
		\begin{prooftree}
		\hypo{ (u_i)_{i\in I} \cohOrred (U'_i)_{i\in I} }
		\infer1[\cohresource*≥0]
			{ (u_i)_{i\in I} \cohOrred[≥0] (U'_i)_{i\in I} }
		\end{prooftree}
	\qquad
		\begin{prooftree}
		\hypo{ (s_i)_{i\in I} \cohOrred[≥d+1] (S'_i)_{i\in I} }
		\infer1[λ_{\cohresource≥d+1}]
			{ (λx.s_i)_{i\in I} \cohOrred[≥d+1] (λx.S'_i)_{i\in I} }
		\end{prooftree}
	\end{widecenter}
	
	\vspace{\topsep}
	\begin{widecenter}
		\begin{prooftree}
		\hypo{ (s_i)_{i\in I} \cohOrred[≥d+1] (S'_i)_{i\in I} }
		\hypo{ \forall i,j,\ \ms t_i \coh \ms t_j }
		\infer2[@l_{\cohresource≥d+1}]{ (\rapp{s_i}t_i)_{i\in I} 
			\cohOrred[≥d+1] (\rapp{\smash{S'_i}}t_i)_{i\in I} }
		\end{prooftree}
	\qquad
		\begin{prooftree}
		\hypo{ \forall i,j,\ s_i \coh s_j  }
		\hypo{ (\ms t_i)_{i\in I} \cohOrred[≥d] (\ms T'_i)_{i\in I} }
		\infer2[@r_{\cohresource≥d+1}]{ (\rapp{s_i}t_i)_{i\in I} 
			\cohOrred[≥d+1] (\rapp{s_i}T'_i)_{i\in I} }
		\end{prooftree}
	\end{widecenter}
	
	\vspace{\topsep}
	\begin{widecenter}
		\begin{prooftree}
			\hypo{ (t_{i,j})_{\begin{subarray}{l}
					i\in I \\ 1\le j\le k_i
				\end{subarray}} \cohOrred[≥d+1] 
				(T'_{i,j})_{\begin{subarray}{l}
					i\in I \\ 1\le j\le k_i
				\end{subarray}} }
			\infer1[!_{\cohresource≥d+1}]
				{ ([ t_{i,1},\dotsc, t_{i,k_i} ])_{i\in I} \cohOrred[≥d+1] 
				([ T'_{i,1},\dotsc, T'_{i,k_i} ])_{i\in I} }
		\end{prooftree}
	\qquad
		\begin{prooftree}
		\hypo{ (u_i)_{i\in I} \cohOrred (U'_i)_{i\in I} }
		\infer1[\Sigma_{\cohresource}]{ \sum_{i \in I} a_i u_i \cohrred 
			\sum_{i \in I} a_i U'_i. }
		\end{prooftree}
	\end{widecenter}
\end{defi}

Without giving all the details of their (rather straightforward) proofs,
let us state two crucial properties of the uniform resource reduction.
The first one means that whenever we analyse a reduction $\UU \cohrred \VV$,
we can choose whichever index set $I$ we like in the backwards
application of the rule \proofrule{\Sigma_{\cohresource}},
which turns out to be extremely helpful!

\begin{lem}[{\cite[Lemma 4.46]{CerdaPhD}}]
\label{lem:cohOrred-choose-index-set}
	For all reduction $(u_i)_{i \in I} \cohOrred (U'_i)_{i \in I}$
	between families $(u_i)_{i \in I} \in (\rc**)^I$ and
	$(U'_i)_{i \in I} \in (\frsums**)^I$,
	for $i,j \in I$,
	if $u_i = u_j$
	then $U'_i = U'_j$.
\end{lem}

The second one (which does in fact crucially use the previous one)
expresses the fact that \cref{defi:cohrred}
does not depart from the previous framework:
the uniform lifting is a strengthening of the usual one.

\begin{lem}[{\cite[Corollary 4.47]{CerdaPhD}}]
	For all $\SS,\TT \in \irsums$, if $\SS \cohrreds \TT$
	then $\SS \Lrred \TT$.
\end{lem}

Before we move on to the simulation and conservativity properties
we are actually interested in,
let us state one more result about the uniform resource reduction,
as we have been using it the proof of \cref{thm:acc}.
Observe how uniformity allows for a very strong control over the reductions,
and therefore a simple proof,
whereas the corresponding statement for $\rreds$ and $\Lrred$
(instead of $\cohrred$ and $\cohrreds$)
turns out to be quite tricky to prove or disprove
--- just as for the transitivity of $\Lrred$,
which remains an open question \cite[Open question~3.17]{CerdaPhD}.

\begin{lem} \label{lem:cohrred-from-sums-to-fsums}
	For all sums $\SS, \TT, \UU \in \irsums$
	such that $\SS \cohrred \TT + \UU$
	and $\sumsupp{\TT} \cap \sumsupp{\UU} = \emptyset$,
	and for all decomposition $\SS = \sum_{i \in I} a_i \cdot S_i$
	into finitely supported sums,
	there are finitely supported sums $T_i, U_i$ (for $i \in I$)
	such that $\TT = \sum_{i \in I} a_i \cdot T_i$,
	$\UU = \sum_{i \in I} a_i \cdot U_i$,
	and for all $i \in I$, $S_i \cohrred T_i + U_i$.
\end{lem}
	
	\begin{proof}
	If we decompose each $S_i$ into $\sum_{j=1}^{n_i} s_{i,j}$,
	the reduction $\SS \cohrred \TT + \UU$ must have been obtained
	by the following application of the rule 
	\proofrule{\Sigma_{\cohresource}}
	(where, as already explained, we can choose a convenient
	index set thanks to \cref{lem:cohOrred-choose-index-set}):
	\[\begin{prooftree}
		\hypo{ (s_{i,j})_{\substack{i \in I \\ 1 \leq j \leq n_i}}
			& \cohOrred (S'_{i,j})_{\substack{i \in I \\ 1 \leq j \leq n_i}} 
			}
		\infer1[\Sigma_{\cohresource}]{ \SS = \sum_{i \in I} 
		\sum_{j=1}^{n_i} 
				a_i \cdot s_{i,j}
			& \cohrred \TT + \UU = \sum_{i \in I} \sum_{j=1}^{n_i} 
				a_i \cdot S'_{i,j}
			}
	\end{prooftree}\]
	Since $\TT$ and $\UU$ have disjoint supports,
	for each $i \in I$ and $1 \leq j \leq n_i$
	we can decompose $S'_{i,j} = T_{i,j} + U_{i,j}$
	with $T_{i,j} \suminclusion \TT$ and $U_{i,j} \suminclusion \UU$.
	Therefore we can conclude with
	$T_i \eqdef \sum_{j=1}^{n_i} T_{i,j}$
	and $U_i \eqdef \sum_{j=1}^{n_i} U_{i,j}$,
	for all $i \in I$.
	\end{proof}

As we are now convinced that we obtained a much better behaved
lifting of the resource reduction to (uniform) sums,
let us see if it is still permissive enough to simulate β-reduction.
In the previous parts of this article, we have been considering
the simulation of finite β-reduction acting on finite λ-terms
(\cref{thm:Lrred-simul-breds} and \cref{sec:mashup}),
and of infinitary β-reduction acting on infinitary λ-terms
(\cref{thm:Lrred-simul-bredi} and \cref{sec:acc});
in the following we only consider infinitary λ$\bot$-terms
but we distinguish finite and infinitary β-reductions.

For the former, it is enough to observe that
all the pointwise reductions $\Lrred$
occurring in the proof of \cref{thm:Lrred-simul-breds}
are in fact instances of the particular case $\cohrreds$,
hence the following reformulation.

\begin{simulthm}[{\cite{CerdaPhD}}, Lemma 4.50]
\label{thm:cohrred-simul-bred}
	For all λ$\bot$-terms $M,N \in \lbci$, if $M \bred[≥d] N$ then
	$\Tay(M) \cohrred[≥d] \Tay(N)$.
\end{simulthm}

However, $\cohrreds$ is not sufficient any more to simulate
the infinitary β-reduction $\bredi$:
we need more than the reflexive-transitive closure.
The simulating reduction we are looking for needs to be:
\begin{itemize}
\item an extension of $\cohrreds$, because we want to be able
	to simulate not only finite, but also infinite sequences of reductions,
\item a restriction of $\Lrred$, because we want to eliminate
	the non-uniform reductions that cannot be turned
	into actual β-reductions,
	in order to obtain a conservativity result in the end.
\end{itemize}
The way we proceed is guided by the stratification property
(\cref{lem:stratification}).

\begin{nota}
	The (applicative) \defemph{depth} of a resource term is the integer
	defined by
	\begin{align*}
		\rdepth(x) &\eqdef 0 &
		\rdepth(\rapp st)
			& \eqdef \max\left( \rdepth(s) , 1+\rdepth(\ms{t}) \right) \\
		\rdepth(\lambda x.s) & \eqdef \rdepth(s) &
		\rdepth([t_1,\dotsc,t_n])
			& \eqdef \max_{1\leq i\leq n} \rdepth(t_i).
	\end{align*}
	Using this notation,
	for all sum $\sum_{i\in I} a_i \cdot s_i \in \irsums$
	and integer $d \in \Nats$ we define:
	\[	\abovedepth*{ \sum_{i\in I} a_i \cdot s_i }
		\eqdef \sum_{\substack{ i\in I \\ \rdepth(s_i) < d }}
		a_i \cdot s_i. \]
\end{nota}

\begin{defi} \label{defi:cohrredi}
	The relation $\mathord{\cohrredi} \subset \irsums** \times \irsums**$
	of \defemph{infinitary uniform resource reduction} is defined
	by writing $\UU \cohrredi \VV$ whenever
	there is a sequence $(\UU_d)_{d \in \Nats}$
	such that
	\[	\UU_0 = \UU \qquad 
		\forall d\in\Nats,\ \UU_d \cohrreds[≥d] \UU_{d+1} \qquad 
		\forall d\in\Nats,\ \abovedepth{\UU_d} = \abovedepth{\VV}. \]
\end{defi}

By design, $\cohrredi$ mimics the stratification
of an infinitary β-reduction, hence the following property.

\begin{simulcor}[of \cref{lem:stratification,thm:cohrred-simul-bred}]
\label{cor:cohrredi-simul-bredi}
	For all $M,N \in \lbci$,
	if $M\bredi N$ then $\Tay(M) \cohrredi \Tay(N)$.
\end{simulcor}
	
	\begin{proof}
	We need to define a sequence $(\UU_d)_{d \in \Nats}$
	as in \cref{defi:cohrredi}.
	By stratification (\cref{lem:stratification}),
	we obtain a sequence $(M_d) \in (\lbci)^{\Nats}$
	such that for all $d\in\Nats$,
	\[	M = M_0 \breds[≥0] M_1 \breds[≥1] M_2 \breds[≥2] \dots
		\breds[≥d-1] M_d \bredi[≥d] N \]
	and we can define $\UU_d \eqdef \Tay(M_d)$.
	The conclusion follows immediately by
	\cref{thm:cohrred-simul-bred} and by the fact that if $M \bredi[≥d] N$,
	then $\abovedepth{\Tay(M)} = \abovedepth{\Tay(N)}$.
	\end{proof}

\subsection{Conservativity \texorpdfstring{\wrt}{wrt.} 
the infinitary λ-calculus}
\label{sec:uniformity:conservativity}

As announced, the simulation stated in \cref{cor:cohrredi-simul-bredi}
enjoys a converse conservativity property:

\begin{conservthm} \label{thm:conserv-cohrredi-bredi}
	For $M,N \in \lbci$,
	if $\Tay(M) \cohrredi \Tay(N)$
	then $M \bredi N$.
\end{conservthm}

The purpose of this subsection is to provide the proof of this theorem.
We first prove the following lemma, which expresses
how we will use uniformity in the main proof.

\begin{lem} \label{lem:conserv-cohrred-bred-improved}
	For all $M \in \lbci$, $\SS \in \irsums$ and $d \in \Nats$,
	if $\Tay(M) \cohrred[≥d] \SS$
	then there exists an $N \in \lbci$
	such that $\SS = \Tay(N)$ and $M \bred[≥d] N$.
\end{lem}
	
	\begin{proof}
	By the backwards application of the rule 
	\proofrule{\Sigma_{\cohresource≥d}},
	there is a family of finite sums $T_s$ such that
	\begin{equation} \label{lem:conserv-cohrred-bred-improved:eq:decomp}
		(s)_{s \in \sumsupp{\Tay(M)}} \cohOrred[≥d] 
			(T_s)_{s \in \sumsupp{\Tay(M)}}
		\qquad \text{and} \quad
		\SS = \sum_{s \in \sumsupp{\Tay(M)}} \Tay(M,s) \cdot T_s.
	\end{equation}
	We proceed by induction on this reduction,
	following the rules of \cref{defi:cohrred}.
	As all the inductive cases are straightforward,
	we concentrate in the base case where $d = 0$ and
	all $s \in \sumsupp{\Tay(M)}$ are of the shape
	$s = (λx.u_s)\ms v_s$ with $T_s = \rsubst*{u_s}{\ms v_s}$.
	By construction of the Taylor expansion $M$ and its approximants
	have the same shape, hence there are $P,Q \in \lbci$
	such that $M = (λx.P)Q$.
	In addition	by \cref{lem:cohOrred-choose-index-set}
	we can restate \cref{lem:conserv-cohrred-bred-improved:eq:decomp}
	as follows:
	\begin{gather*}
		\left((λx.u)\ms v\right)_{\substack{
			u \in \sumsupp{\Tay(P)} \\ \ms v \in \sumsupp{\Tay(Q)^!}
			}}
			\cohOrred \left(\rsubst uv\right)_{\substack{
			u \in \sumsupp{\Tay(P)} \\ \ms v \in \sumsupp{\Tay(Q)^!}
			}},
		\\
		\shortintertext{and:}
		\SS = \sum_{\substack{
			u \in \sumsupp{\Tay(P)} \\ \ms v \in \sumsupp{\Tay(Q)^!}
			}} \Tay(M, (λx.u)\ms v) \cdot \rsubst uv.
	\end{gather*}
	Then we can apply the classical simulation of the substitution 
	by the Taylor expansion, see \eg \cite[Lemma~4.8]{Vaux19},
	obtaining:
	\[	\Tay(\subst PQ) = \rsubst* {\Tay(P)} {\Tay(Q)^!}
		= \sum_{\substack{
			u \in \sumsupp{\Tay(P)} \\ \ms v \in \sumsupp{\Tay(Q)^!}
			}} \Tay(P,u) \times \Tay^!(Q,\ms v) \cdot \rsubst uv 
		= \SS \]
	therefore we can conclude with $N \eqdef \subst PQ$.
	\end{proof}

Observe that the conservativity of the simulation of $(\lbci,\breds)$
by $(\irsums,\cohrreds)$ appears as a particular case of this lemma:

\begin{conservcor}\label{cor:conserv-cohrred-bred}%
	\scalebox{.975}[1]{%
	For all $M,N \in \lbci$, if $\Tay(M) \cohrred \Tay(N)$
	then $M \bred N$.%
	}
\end{conservcor}

This observation being made, we are ready for the proof of conservativity
for the infinitary simulation.

\begin{proof}[Proof of theorem \ref{thm:conserv-cohrredi-bredi}]
	Suppose that there is a sequence $(\SS_d)_{d \in \Nats}$
	such that
	\[	\SS_0 = \Tay(M) \qquad 
		\forall d\in\Nats,\ \SS_d \cohrreds[≥d] \SS_{d+1} \qquad 
		\forall d\in\Nats,\ \abovedepth{\SS_d} = \abovedepth{\Tay(N)}. \]
	By iterated applications of \cref{lem:conserv-cohrred-bred-improved}
	to the first two hypotheses,
	there is a sequence of terms $(M_d)_{d \in \Nats}$, with $M_0 = M$,
	such that $\forall d \in \Nats,\ \SS_d = \Tay(M_d)$,
	as well as:
	\begin{gather}
		\forall d\in\Nats,\ M_d \breds[≥d] M_{d+1},
			\label{thm:conserv-cohrredi-bredi:h1} \\
	\shortintertext{whereas the third hypothesis becomes:}
		\forall d\in\Nats,\ \abovedepth{\Tay(M_d)} 
			= \abovedepth{\Tay(N)}.
			\label{thm:conserv-cohrredi-bredi:h2}
	\end{gather}
	For any sequence $(M_d)_{d \in \Nats}$ such that
	\cref{thm:conserv-cohrredi-bredi:h1,thm:conserv-cohrredi-bredi:h2}
	hold, we build a reduction $M_0 \bredi N$
	by nested induction and coinduction on $N$.
	
	\begin{itemize}
	\item Case $N = x$. 
		By \cref{thm:conserv-cohrredi-bredi:h2},
		\[ \abovedepth[1]{\Tay(M_1)} = \abovedepth[1]{\Tay(N)} = x \]
		hence also $\Tay(M_1) = x$, and finally $M_1 = x$.
		By applying the rule \proofrule{\Vars_β^{001}}
		from \cref{defi:bredi}
		to $M \breds x$, we obtain $M \bredi x = N$.
	
	\item Case $N = \bot$. 
		By \cref{thm:conserv-cohrredi-bredi:h2},
		\[ \abovedepth[1]{\Tay(M_1)} = \abovedepth[1]{\Tay(N)} = 0 \]
		hence also $\Tay(M_1) = 0$, and finally $M_1 = \bot$.
		By applying the rule \proofrule{\bot_β^{001}}
		from \cref{defi:bredi}
		to $M \breds \bot$, we obtain $M \bredi \bot = N$.
	
	\item Case $N = λx.P'$. 
		By \cref{thm:conserv-cohrredi-bredi:h2}, for all $d≥1$,
		\[	\abovedepth{\Tay(M_d)} = \abovedepth{\Tay(N)} = 
			λx.\abovedepth{\Tay(P')} \]
		hence there is a term $P_d \in \lci$ such that
		$M_d = λx.P_d$ verifying:
		\begin{itemize}
		\item that $P_d \breds[≥d] P_{d+1}$,
			by the rule \proofrule{λ_{β≥d}}
			from \cref{lem:stratification} applied to
			\cref{thm:conserv-cohrredi-bredi:h1},
		\item that $\abovedepth{\Tay(P_d)} = \abovedepth{\Tay(P')}$,
			by \cref{thm:conserv-cohrredi-bredi:h2}.
		\end{itemize}
		We also define $P_0 \eqdef P_1$,
		so that the sequence $(P_d)_{d\in\Nats}$ satisfies
		\cref{thm:conserv-cohrredi-bredi:h1,%
		thm:conserv-cohrredi-bredi:h2} \wrt~$P'$.
		By induction we can build a reduction $P_0 = P_1 \bredi P'$.
		Since $M_0 \breds M_1 = λx.P_1$,
		we can apply the rule \proofrule{λ_β^{001}} from \cref{defi:bredi}
		and obtain $M_0 \bredi λx.P' = N$.
	
	\item Case $N = (P')Q'$. 
		By \cref{thm:conserv-cohrredi-bredi:h2}, for all $d≥1$,
		\[	\abovedepth{\Tay(M_d)} = \abovedepth{\Tay(N)} = 
			\left(\abovedepth{\Tay(P')}\right) 
				\abovedepth[d-1]{\Tay(Q')^!} \]
		hence there are terms $P_d, Q_d \in \lci$ such that
		$M_d = (P_d)Q_d$ verifying:
		\begin{itemize}
		\item that $P_d \breds[≥d] P_{d+1}$
			and $Q_d \breds[≥d-1] Q_{d+1}$,
			by the rule \proofrule{@_{β≥d}}
			from \cref{lem:stratification} applied to
			\cref{thm:conserv-cohrredi-bredi:h1},
		\item that $\abovedepth{\Tay(P_d)} = \abovedepth{\Tay(P')}$
			and $\abovedepth[d-1]{\Tay(Q_d)} = \abovedepth[d-1]{\Tay(Q')}$,
			by \cref{thm:conserv-cohrredi-bredi:h2};
			for the second equality one also uses the fact that
			$\abovedepth{\SS^!} = \abovedepth{\TT^!}$
			implies $\abovedepth \SS = \abovedepth \TT$,
			which is straightforward.
		\end{itemize}
		We also define $P_0 \eqdef P_1$,
		so that the sequences $(P_d)_{d\in\Nats}$
		and $(Q_{d+1})_{d\in\Nats}$ satisfy
		\cref{thm:conserv-cohrredi-bredi:h1,%
		thm:conserv-cohrredi-bredi:h2},
		respectively \wrt~$P'$ and $Q'$.
		Respectively by induction and coinduction,
		we can build reductions $P_0 = P_1 \bredi P'$
		and $Q_1 \bredi Q'$.
		Since $M_0 \breds M_1 = (P_1)Q_1$,
		we can apply the rule \proofrule{@_β^{001}} from \cref{defi:bredi}
		and obtain $M_0 \bredi (P')Q' = N$.
	\qedhere
	\end{itemize}
\end{proof}

In particular, observe that there is no reduction $\acc \cohrredi \acc*$:
in the sequence of reductions given in \cref{eq:acc:proof-it1:1}
in the proof of \cref{lemma:acc-i}
(that is to say the proof of item (i) of \cref{thm:acc}),
all steps $\Tay(\acc_{d}) \Lrred \Tay(\acc_{d+1})$
can be turned into $\Tay(\acc_{d}) \cohrreds \Tay(\acc_{d+1})$
as a consequence of the uniform simulation \cref{thm:cohrred-simul-bred},
but not into $\Tay(\acc_{d}) \cohrreds[≥d] \Tay(\acc_{d+1})$:
indeed, there is always a reduction step occurring at depth~0.

\subsection{An epilogue on β\texorpdfstring{$\bot$}{⊥}-reductions} 
\label{sec:uniformity:beta-bottom}

In all the above developments, we did not take into account
the $\bot$-reduction steps introduced in \cref{defi:bbotred}.
However, these steps are needed in order
to ensure the confluence of the infinitary λ-calculus \cite{KennawayEtAl97}
and in particuler to express the reductions $M \red[β\bot][001] \BT(M)$,
which are the main motivation for considering the (001-)infinitary λ-calculus
and lie at the core of Ehrhard and Regnier's commutation theorem.
Fortunately, there is almost nothing to add to the previous work
in order to extend it to β$\bot$-reduction;
let us work this out before concluding.

First, let us recall from \cref{cor:Lrred-simul-bbotredi}
that $(\irsums, \Lrred)$ also simulates $(\lbci, \bbotredi)$,
and make the following observation:
the Accordion is also a counter-example to the conservativity
of this simulation.
This can be deduced from the following lemma:

\begin{lem} \label{lem:bbotred-to-lci-is-bred}
	For all $M \in \lbci$ and $N \in \lci$,
	if $M \bbotredi N$ then $M \bredi N$.
\end{lem}
	
	\begin{proof}
	Denote by $\botred$ the reduction obtained
	by ommitting the base case \proofrule{beta_{β\bot}}
	in \cref{defi:bbotred}
	(\ie by only considering \proofrule{erase_{\bot}} as a base case),
	and by $\botredi$ its 001-infinitary closure by \cref{defi:bredi}.
	A standard result in infinitary λ-calculus states that
	whenever $M \bbotredi N$, there exists a term $N'$
	such that $M \bredi N' \botredi N$
	(the original proof is \cite[Lemma~46(iii)]{KennawayEtAl97},
	but it can also be seen as a consequence of the observation that
	$M \botred M' \bred N$ implies the existence of $M''$
	such that $M \bred M'' \botreds N$).
	In this situation, if $N \in \lci$ then there is no other choice
	than $N' = N$, since any $\bot$-reduction step would introduce
	occurrences of $\bot$ in $N$.
	\end{proof}

By contraposition, since $\acc* \in \lci$,
the inexistence of a reduction $\acc \bredi \acc*$ (\cref{thm:acc})
ensures that there is no reduction $\acc \bbotredi \acc*$ either.
As a consequence, the motivation of trying to build
a conservative simulation using a uniform lifting of $\rred$
is also applicable to β$\bot$-reductions.
The additional ingredient is as follows.

\begin{defi}
	Given an index set $I$, an extended auxiliary relation
	$\mathord{\cohOrbotred} \subset (\rc**)^I \times (\frsums**)^I$
	is defined by the rules \proofrule{beta_{\cohresource}}
	to \proofrule{!_{\cohresource}} from \cref{defi:cohrred},
	together with the following rule:
	\begin{prooftree*}
	\hypo{ \forall i,j,\ u_i \coh u_j }
	\hypo{ \forall i,\ u_i \rreds 0 }
	\infer2[erase_{\cohresource}]
		{ (u_i)_{i \in I} \cohOrbotred (0)_{i \in I}. }
	\end{prooftree*}
	Relations $\mathord{\cohrbotred}, \mathord{\cohrbotred[≥d]}
	\subset \irsums** \times \irsums**$, for $d \in \Nats$,
	are defined from $\cohOrbotred$ just as in
	the remainder of \cref{defi:cohrred}.
\end{defi}

This definition is justified by a well-known characterisation
(see \cite[Theorem~5.6]{CerdaVauxAuclair23}):
a λ$\bot$-term $M$ has no head normal form
iff all its resource approximants can be reduced to~0;
in other terms, $M \bbotred \bot$ iff $\Tay(M) \Lrred 0$.
Therefore we extended the uniform resource reduction
just as we needed in order to simulate β$\bot$-reduction:

\begin{simulthm}[{\cite{Cerda25}, Theorem 4.52}]
\label{thm:cohrbotred-simul-bbotred}
	For all $M,N \in \lbci$,
	if $M \bbotred[≥d] N$
	then $\Tay(M) \cohrbotred[≥d] \Tay(N)$.
\end{simulthm}

As for simulating infinitary β$\bot$-reduction,
there is nothing to change to the method introduced
in \cref{sec:uniformity:definitions}.

\begin{defi}
	A relation $\mathord{\cohrbotredi} \subset \irsums** \times \irsums**$
	is defined from $\cohrbotred$ exactly as in \cref{defi:cohrredi},
	\ie by writing $\UU \cohrbotredi \VV$ whenever
	there is a sequence $(\UU_d)_{d \in \Nats}$
	such that
	\[	\UU_0 = \UU \qquad 
		\forall d\in\Nats,\ \UU_d \cohrbotreds[≥d] \UU_{d+1} \qquad 
		\forall d\in\Nats,\ \abovedepth{\UU_d} = \abovedepth{\VV}. \]
\end{defi}

\begin{simulcor}
\label{cor:cohrbotredi-simul-bbotredi}
	For all $M,N \in \lbci$,
	if $M \bbotredi N$
	then $\Tay(M) \cohrbotredi \Tay(N)$.
\end{simulcor}
	
	\begin{proof}
	Just as we did for proving \cref{thm:cohrred-simul-bred},
	the result is a straightforward translation of
	\cref{lem:stratification} into the resource calculus,
	using \cref{thm:cohrbotred-simul-bbotred}.
	\end{proof}

Finally, let us state and prove that these two simulations
are again conservative, which concludes our exposition
of the benefits of uniformity.

\begin{conservthm} \label{thm:conserv-cohrbotred-botred}
	For all $M,N \in \lbci$,
	if $\Tay(M) \cohrbotred[≥d] \Tay(N)$
	then $M \bbotred[≥d] N$.
\end{conservthm}
	
	\begin{proof}
	As we did in \cref{lem:conserv-cohrred-bred-improved},
	we prove a more general result:
	for all $M \in \lbci$, $\SS \in \irsums$ and $d \in \Nats$,
	if $\Tay(M) \cohrbotred[≥d] \SS$
	then there exists an $N \in \lbci$
	such that $\SS = \Tay(N)$ and $M \bbotred[≥d] N$.
	The proof is identical, with the difference that 
	the induction has one more case, 
	corresponding to the rule \proofrule{erase_{\cohresource}}:
	in this case, $\forall s \in \sumsupp{\Tay(M)},\ s \rreds 0$
	and $\SS = 0$.
	by the above mentionned characterisation
	\cite[Theorem~5.6]{CerdaVauxAuclair23},
	the first hypothesis means that $M \bbotred \bot$,
	while the second one is equivalent to $\SS = \Tay(\bot)$,
	hence we can take $N \eqdef \bot$.
	\end{proof}

\begin{conservthm} \label{thm:conserv-cohrbotredi-botredi}
	For all $M,N \in \lbci$,
	if $\Tay(M) \cohrbotredi \Tay(N)$
	then \mbox{$M \bbotredi N$}.
\end{conservthm}
	
	\begin{proof}
	The proof is exactly identical to 
	the proof \cref{thm:conserv-cohrredi-bredi}:
	just as the latter used \cref{lem:conserv-cohrred-bred-improved}
	as a blackbox,
	we use its counterpart that we just established in the proof of
	\cref{thm:conserv-cohrbotred-botred}.
	\end{proof}

%*******************************************************************************

\section{Summary and conclusive remarks} \label{sec:conclusion}

\colorlet{cbase}{blue!20!gray!50}
\colorlet{cres}{cyan!30!gray!40}
\colorlet{cout}{black!10}
\colorlet{cyes}{green!70!black!60}
\colorlet{cno}{red!50}

\tikzstyle{node} = [fill=cbase]
\tikzstyle{nsim} = [node,draw]
\tikzstyle{nres} = [fill=cres]
\tikzstyle{nout} = [fill=cout, text=cout!50!black]
\tikzstyle{nyes} = [fill=cyes]
\tikzstyle{nno} = [fill=cno]

\tikzstyle{l} = [cbase]
\tikzstyle{lr} = [cres]
\tikzstyle{lnc} = [cbase, dashed]
\tikzstyle{lout} = [cout]
\tikzstyle{lyes} = [cyes]
\tikzstyle{lno} = [cno]

\tikzstyle{tsim} = [scale=.7]
\tikzstyle{tyes} = [cyes!80!black]
\tikzstyle{tno} = [cno!80!black]

\paragraph{A \enquote{simulation poset} of reduction systems}

In this paper, we have been considering
the linear approximation of the λ-calculus
in the light of the simulations it induces
between various reduction systems.
Indeed, this linear approximation translates λ-terms
into weighted sum of approximants taken
in a multilinear λ-calculus, the resource λ-calculus,
\emph{via} an operation of Taylor expansion,
in such a way that the sums of approximants are endowed with
a reduction relation simulating the reduction of the approximated terms.
More precisely, we considered several simulated λ-calculi:
\begin{itemize}
\item where the λ-terms may or may not be extended with 
	an \enquote{undefined} constant $\bot$,
\item where the λ($\bot$)-terms may be finite or 001-infinitary,
\item where the β-reduction may or may not be extended
	into a β$\bot$-reduction collapsing unsolvable
	(\ie non-head-normalising) terms to $\bot$,
\item where the β($\bot$)-reduction may be the usual, finite one,
	or its 001-infinitary closure;
\end{itemize}
as well as several simulating reductions acting on sums of resource terms:
\begin{itemize}
\item the usual lifting to sums of the resource reduction, $\Lrred$,
\item several variants of a \emph{coherent} lifting that we introduce.
\end{itemize}
All these possibilites give rise to many reduction systems,
that we summarise in the diagram below.

\begin{figure}[H]
\begin{center}\footnotesize
\begin{tikzpicture}[x=2cm]
\node[node] (l-bs) 		at (2,4) {$(\lc, 	\breds)$};
\node[node] (lb-bs) 	at (1,3) {$(\lbc, 	\breds)$};
\node[node] (li-bs) 	at (3,3) {$(\lci, 	\breds)$};
\node[node] (lb-bbs) 	at (0,2) {$(\lbc, 	\bbotreds)$};
\node[node] (lbi-bs) 	at (2,2) {$(\lbci, 	\breds)$};
\node[node] (li-bi) 	at (4,2) {$(\lci, 	\bredi)$};
\node[node] (lbi-bbs) 	at (1,1) {$(\lbci, 	\bbotreds)$};
\node[node] (lbi-bi) 	at (3,1) {$(\lbci, 	\bredi)$};
\node[node] (lbi-bbi) 	at (2,0) {$(\lbci, 	\bbotredi)$};
\draw[l]	(l-bs) 		-- 				(lb-bs);
\draw[l]	(l-bs) 		-- 				(li-bs);
\draw[l]	(lb-bs) 	-- 				(lbi-bs);
\draw[l]	(li-bs) 	-- 				(lbi-bs);
\draw[lnc]	(lb-bs) 	-- 				(lb-bbs);
\draw[l]	(lb-bbs) 	-- 				(lbi-bbs);
\draw[lnc]	(lbi-bs) 	-- 				(lbi-bbs);
\draw[lnc]	(li-bs) 	-- 				(li-bi);
\draw[lnc]	(lbi-bs) 	-- 				(lbi-bi);
\draw[l]	(li-bi) 	-- 				(lbi-bi);
\draw[lnc]	(lbi-bbs) 	-- 				(lbi-bbi);
\draw[lnc]	(lbi-bi) 	-- 				(lbi-bbi);
\draw[l]	(l-bs) 		to [bend right]	(lb-bbs);
\draw[l]	(l-bs) 		to [bend left]	(li-bi);
\draw[l]	(lb-bbs)	to [bend left]	(lbi-bbi);
\draw[l]	(li-bi)		to [bend right]	(lbi-bbi);
\draw[l]	(lb-bs)		to [bend left]	(lbi-bi);
\draw[l]	(li-bs)		to [bend right]	(lbi-bbs);
\node[nres] (r-cs)		at (3,-1) {$(\irsums, \cohrreds)$};
\node[nres] (r-cbs)		at (2,-2) {$(\irsums, \cohrbotreds)$};
\node[nres] (r-ci)		at (4,-2) {$(\irsums, \cohrredi)$};
\node[nres] (r-cbi)		at (3,-3) {$(\irsums, \cohrbotredi)$};
\node[nres] (r-L)		at (3,-4.5) {$(\irsums, \Lrred)$};
\draw[lnc,lr]	(r-cs)		to				 (r-cbs);
\draw[lnc,lr]	(r-cs)		to				 (r-ci);
\draw[lnc,lr]	(r-cbs)		to				 (r-cbi);
\draw[lnc,lr]	(r-ci)		to				 (r-cbi);
\draw[lr]		(lbi-bs)	to	[] (r-cs);
\draw[lr]		(lbi-bbs)	to	[] (r-cbs);
\draw[lr]		(lbi-bi)	to	[ ] (r-ci);
\draw[lr]		(lbi-bbi)	to	[ ] (r-cbi);
\draw[lnc,lr]	(r-cbi)		to				 (r-L);
\end{tikzpicture}
\end{center}
\end{figure}

In this diagram, each link from a reduction system
down to another reduction system means that
the latter simulates the former.
It is an immediate consequence of \cref{defi:extension}
that simulation is transitive,
and induces an order on reduction systems.

In addition, these links are solid when the simulation is conservative,
and dashed when it is non-conservative.
The following table justifies the (non-)conservativity of
the simulations between all the λ($\bot$)-calculi we consider.

\begin{table}[H]
\begin{center}\small
\begin{tabular}{ccccccc}
	\toprule
	\textbf{Simulations} &
		\adjustbox{valign=c}{\begin{tikzpicture}[x=2ex,y=1.5ex]
		\draw		(2,4) 	-- 				(1,3);
		\draw[l]	(2,4) 	-- 				(3,3);
		\draw[l]	(1,3) 	-- 				(2,2);
		\draw		(3,3) 	-- 				(2,2);
		\draw[l]	(1,3) 	-- 				(0,2);
		\draw[l]	(0,2) 	-- 				(1,1);
		\draw[l]	(2,2) 	-- 				(1,1);
		\draw[l]	(3,3) 	-- 				(4,2);
		\draw[l]	(2,2) 	-- 				(3,1);
		\draw		(4,2) 	-- 				(3,1);
		\draw[l]	(1,1) 	-- 				(2,0);
		\draw[l]	(3,1) 	-- 				(2,0);
		\draw[l]	(2,4) 	to [bend right]	(0,2);
		\draw[l]	(2,4) 	to [bend left]	(4,2);
		\draw[l]	(0,2)	to [bend left]	(2,0);
		\draw[l]	(4,2)	to [bend right]	(2,0);
		\draw[l]	(1,3)	to [bend left]	(3,1);
		\draw[l]	(3,3)	to [bend right]	(1,1);
		\end{tikzpicture}}
	&
		\adjustbox{valign=c}{\begin{tikzpicture}[x=2ex,y=1.5ex]
		\draw[l]	(2,4) 	-- 				(1,3);
		\draw[l]	(2,4) 	-- 				(3,3);
		\draw[l]	(1,3) 	-- 				(2,2);
		\draw[l]	(3,3) 	-- 				(2,2);
		\draw[l]	(1,3) 	-- 				(0,2);
		\draw[l]	(0,2) 	-- 				(1,1);
		\draw[l]	(2,2) 	-- 				(1,1);
		\draw[l]	(3,3) 	-- 				(4,2);
		\draw[l]	(2,2) 	-- 				(3,1);
		\draw[l]	(4,2) 	-- 				(3,1);
		\draw[l]	(1,1) 	-- 				(2,0);
		\draw[l]	(3,1) 	-- 				(2,0);
		\draw		(2,4) 	to [bend right]	(0,2);
		\draw[l]	(2,4) 	to [bend left]	(4,2);
		\draw[l]	(0,2)	to [bend left]	(2,0);
		\draw		(4,2)	to [bend right]	(2,0);
		\draw[l]	(1,3)	to [bend left]	(3,1);
		\draw		(3,3)	to [bend right]	(1,1);
		\end{tikzpicture}}
	&
		\adjustbox{valign=c}{\begin{tikzpicture}[x=2ex,y=1.5ex]
		\draw[l]	(2,4) 	-- 				(1,3);
		\draw		(2,4) 	-- 				(3,3);
		\draw		(1,3) 	-- 				(2,2);
		\draw[l]	(3,3) 	-- 				(2,2);
		\draw[l]	(1,3) 	-- 				(0,2);
		\draw		(0,2) 	-- 				(1,1);
		\draw[l]	(2,2) 	-- 				(1,1);
		\draw[l]	(3,3) 	-- 				(4,2);
		\draw[l]	(2,2) 	-- 				(3,1);
		\draw[l]	(4,2) 	-- 				(3,1);
		\draw[l]	(1,1) 	-- 				(2,0);
		\draw[l]	(3,1) 	-- 				(2,0);
		\draw[l]	(2,4) 	to [bend right]	(0,2);
		\draw[l]	(2,4) 	to [bend left]	(4,2);
		\draw[l]	(0,2)	to [bend left]	(2,0);
		\draw[l]	(4,2)	to [bend right]	(2,0);
		\draw[l]	(1,3)	to [bend left]	(3,1);
		\draw[l]	(3,3)	to [bend right]	(1,1);
		\end{tikzpicture}}
	&
		\adjustbox{valign=c}{\begin{tikzpicture}[x=2ex,y=1.5ex]
		\draw[l]	(2,4) 	-- 				(1,3);
		\draw[l]	(2,4) 	-- 				(3,3);
		\draw[l]	(1,3) 	-- 				(2,2);
		\draw[l]	(3,3) 	-- 				(2,2);
		\draw[l]	(1,3) 	-- 				(0,2);
		\draw[l]	(0,2) 	-- 				(1,1);
		\draw[l]	(2,2) 	-- 				(1,1);
		\draw[l]	(3,3) 	-- 				(4,2);
		\draw[l]	(2,2) 	-- 				(3,1);
		\draw[l]	(4,2) 	-- 				(3,1);
		\draw[l]	(1,1) 	-- 				(2,0);
		\draw[l]	(3,1) 	-- 				(2,0);
		\draw[l]	(2,4) 	to [bend right]	(0,2);
		\draw		(2,4) 	to [bend left]	(4,2);
		\draw		(0,2)	to [bend left]	(2,0);
		\draw[l]	(4,2)	to [bend right]	(2,0);
		\draw		(1,3)	to [bend left]	(3,1);
		\draw[l]	(3,3)	to [bend right]	(1,1);
		\end{tikzpicture}}
	&
		\adjustbox{valign=c}{\begin{tikzpicture}[x=2ex,y=1.5ex]
		\draw[l]	(2,4) 	-- 				(1,3);
		\draw[l]	(2,4) 	-- 				(3,3);
		\draw[l]	(1,3) 	-- 				(2,2);
		\draw[l]	(3,3) 	-- 				(2,2);
		\draw		(1,3) 	-- 				(0,2);
		\draw[l]	(0,2) 	-- 				(1,1);
		\draw		(2,2) 	-- 				(1,1);
		\draw[l]	(3,3) 	-- 				(4,2);
		\draw[l]	(2,2) 	-- 				(3,1);
		\draw[l]	(4,2) 	-- 				(3,1);
		\draw[l]	(1,1) 	-- 				(2,0);
		\draw		(3,1) 	-- 				(2,0);
		\draw[l]	(2,4) 	to [bend right]	(0,2);
		\draw[l]	(2,4) 	to [bend left]	(4,2);
		\draw[l]	(0,2)	to [bend left]	(2,0);
		\draw[l]	(4,2)	to [bend right]	(2,0);
		\draw[l]	(1,3)	to [bend left]	(3,1);
		\draw[l]	(3,3)	to [bend right]	(1,1);
		\end{tikzpicture}}
	&
		\adjustbox{valign=c}{\begin{tikzpicture}[x=2ex,y=1.5ex]
		\draw[l]	(2,4) 	-- 				(1,3);
		\draw[l]	(2,4) 	-- 				(3,3);
		\draw[l]	(1,3) 	-- 				(2,2);
		\draw[l]	(3,3) 	-- 				(2,2);
		\draw[l]	(1,3) 	-- 				(0,2);
		\draw[l]	(0,2) 	-- 				(1,1);
		\draw[l]	(2,2) 	-- 				(1,1);
		\draw		(3,3) 	-- 				(4,2);
		\draw		(2,2) 	-- 				(3,1);
		\draw[l]	(4,2) 	-- 				(3,1);
		\draw		(1,1) 	-- 				(2,0);
		\draw[l]	(3,1) 	-- 				(2,0);
		\draw[l]	(2,4) 	to [bend right]	(0,2);
		\draw[l]	(2,4) 	to [bend left]	(4,2);
		\draw[l]	(0,2)	to [bend left]	(2,0);
		\draw[l]	(4,2)	to [bend right]	(2,0);
		\draw[l]	(1,3)	to [bend left]	(3,1);
		\draw[l]	(3,3)	to [bend right]	(1,1);
		\end{tikzpicture}}
	\\ \midrule
	\textbf{Conservative?} & Yes & Yes & Yes & Yes & No & No
	\\
	\textbf{Why?}
		& Trivial & \Cref{lem:bbotred-to-lci-is-bred}
		& Trivial & Easy\footnotemark
		& $\deltadelta \bbotred \bot$ & $(\yfp)x \bredi \infrapp x$
	\\ \bottomrule
\end{tabular}
\end{center}
\end{table}
\footnotetext{If $M \redi[\xi] N$ but $N$ is finite,
	one can show by a direct induction over $N$ that $M \reds[\xi] N$.
	Alternatively, one can apply \cref{lem:stratification}
	and take $d$ to be the applicative depth of $N$.}

\noindent As concerns the remainder of the diagram:
\begin{itemize}
\item each of the four coherent reductions we define on $\irsums$
	conservatively simulates one of the λ($\bot$)-calculi:
	this is the content of our \cref{sec:uniformity};
\item the simulations between these four reductions are all non-conservative,
	the counter-examples being the same as in the simulated 
	λ($\bot$)-calculi,
	namely $\Tay(\deltadelta) \cohrbotred \Tay(\bot)$
	and $\Tay((\yfp)x) \cohrredi \Tay(\infrapp x)$;
\item finally, $(\irsums,\Lrred)$ simulates $(\irsums,\cohrbotredi)$
	non-conservatively, as a consequence of our \cref{thm:acc}:
	$\Tay(\acc) \Lrred \Tay(\acc*)$ but there is no reduction
	$\Tay(\acc) \cohrbotredi \Tay(\acc*)$\footnote{%
		Otherwise by \cref{thm:conserv-cohrbotredi-botredi}
		there would be a reduction $\acc \bbotredi \acc*$,
		hence by \cref{lem:bbotred-to-lci-is-bred}
		a reduction $\acc \bredi \acc*$,
		which is forbidden by \cref{thm:acc}.
		This is an application of
		\cref{obs:green-solid-green,obs:red-solid-red} below.
	}.
\end{itemize}

\paragraph{(Non-)conservativity results for the usual linear approximation}

In an ongoing line of work
\cite{Vaux17,Vaux19,CerdaVauxAuclair23,CerdaPhD},
we reformulated Ehrhard and Regnier's linear approximation of the λ-calculus
as a simulation of the β-reduction by $(\irsums,\Lrred)$.
In \cref{sec:mashup,sec:acc} of this paper
we investigated the conservativity
of this simulation and obtained two significant results:
\begin{itemize}
\item in \cref{thm:conserv-Lrred-breds} we state that
	the linear approximation is conservative 
	\wrt the finite λ-calculus $(\lc,\breds)$,
	which we prove \emph{via} the \enquote{mashup} technique
	from \cite{KerinecVauxAuclair23};
\item in \cref{thm:acc} we state that the linear approximation
	is not conservative \wrt the infinitary λ-calculus $(\lci,\bredi)$,
	which we prove by exhibiting a counter-example,	the Accordion.
\end{itemize}

This answers to the conservativity problem
for only two out of the nine λ($\bot$)-calculi
simulated by $(\irsums,\Lrred)$;
however the picture can be (almost) completed,
as described in the following diagram.
It should be read as follows:
\begin{itemize}
\item the \fbox{black boxes} and their annotations point to
	the result of simulation by $(\irsums,\Lrred)$
	for the given λ($\bot$)-calculus
	(we mention all the results explicitely stated in the paper,
	although in fact all nine simulations are implied
	by the bottom one, by transitivity),
\item the λ($\bot$)-calculus simulated by $(\irsums,\Lrred)$
	in a conservative way is put in a \colorbox{cyes}{green box}
	that is {\color{cyes!80!black} annotated}
	with a pointer to the conservativity result,
\item the λ($\bot$)-calculi simulated by $(\irsums,\Lrred)$
	in a non-conservative way are put in a \colorbox{cno}{red box}
	that may be {\color{cno!80!black} annotated}
	with a pointer to the proof of non-conservativity
	or with a counter-example,
\item one simulation remains \colorbox{cbase}{indeterminate},
	which is discussed in \cref{conj:conserv-Lrred-lbc-bbotreds} below.
\end{itemize}

\begin{figure}[H] \label{fig:simulations-by-Lrred}
\begin{center}\footnotesize
\begin{tikzpicture}[x=2.5cm,y=1.2cm]
\node[nsim,nyes] 	(l-bs) 		at (2,4) {$(\lc, 	\breds)$};
\node[node,nno] 	(lb-bs) 	at (1,3) {$(\lbc, 	\breds)$};
\node[node,nno] 	(li-bs) 	at (3,3) {$(\lci, 	\breds)$};
\node[node] 		(lb-bbs) 	at (0,2) {$(\lbc, 	\bbotreds)$};
\node[node,nno] 	(lbi-bs) 	at (2,2) {$(\lbci, 	\breds)$};
\node[nsim,nno] 	(li-bi) 	at (4,2) {$(\lci, 	\bredi)$};
\node[node,nno] 	(lbi-bbs) 	at (1,1) {$(\lbci, 	\bbotreds)$};
\node[node,nno] 	(lbi-bi) 	at (3,1) {$(\lbci, 	\bredi)$};
\node[nsim,nno] 	(lbi-bbi) 	at (2,0) {$(\lbci, 	\bbotredi)$};
\draw[l]		(l-bs) 		-- 				(lb-bs);
\draw[l]		(l-bs) 		-- 				(li-bs);
\draw[l,lno]	(lb-bs) 	-- 				(lbi-bs);
\draw[l,lno]	(li-bs) 	-- 				(lbi-bs);
\draw[lnc]		(lb-bs) 	-- 				(lb-bbs);
\draw[l]		(lb-bbs) 	-- 				(lbi-bbs);
\draw[lnc]		(lbi-bs) 	-- 				(lbi-bbs);
\draw[lnc]		(li-bs) 	-- 				(li-bi);
\draw[lnc]		(lbi-bs) 	-- 				(lbi-bi);
\draw[l,lno]	(li-bi) 	-- 				(lbi-bi);
\draw[lnc]		(lbi-bbs) 	-- 				(lbi-bbi);
\draw[lnc]		(lbi-bi) 	-- 				(lbi-bbi);
\draw[l]		(l-bs) 		to [bend right]	(lb-bbs);
\draw[l]		(l-bs) 		to [bend left]	(li-bi);
\draw[l]		(lb-bbs)	to [bend left]	(lbi-bbi);
\draw[l,lno]	(li-bi)		to [bend right]	(lbi-bbi);
\draw[l,lno]	(lb-bs)		to [bend left]	(lbi-bi);
\draw[l,lno]	(li-bs)		to [bend right]	(lbi-bbs);
\node[below=0 of l-bs, tyes]
	{\cref{thm:conserv-Lrred-breds}};
\node[above=0 of l-bs, tsim]
	{\cref{thm:Lrred-simul-breds} \cite{Vaux19}};
\node[below=0 of li-bi, tno] 	{\cref{thm:acc}};
\node[above=0 of li-bi, tsim]	{\parbox{3cm}{\centering
	\cref{thm:Lrred-simul-bredi} \\ 
	\cite{CerdaVauxAuclair23,CerdaPhD}
	}};
\node[above=0 of lbi-bbi, tsim] {\parbox{3cm}{\centering
	\cref{cor:Lrred-simul-bbotredi} \\
	\cite{CerdaVauxAuclair23,CerdaPhD}
	}};
\node[below=0 of lb-bs, tno] 	{$\Tay(\Omega) \Lrred \Tay(\bot)$};
\node[below=0 of li-bs, tno] 	{$\Tay((\yfp) x) \Lrred \Tay(\infrapp x)$};
\end{tikzpicture}
\end{center}
\end{figure}

In addition, the diagram contains some (non-)conservativity statements
without annotation. This means that they can be deduced
from other statements using the following observation:

\begin{obs} \label{obs:green-solid-green}
	If $(C, \red[C])$ simulates $(B, \red[B])$ conservatively
	and $(B, \red[B])$ simulates $(A, \red[A])$ conservatively,
	then $(C, \red[C])$ simulates $(A, \red[A])$ conservatively.
	\emph{In the diagram: if \colorbox{cyes}{$\Lambda_2$}
	is below $\Lambda_1$ and they are related with a 
	{\color{cyes}\underline{\normalcolor solid link}},
	then \colorbox{cyes}{$\Lambda_1$}.}
\end{obs}

\noindent or, in fact, its contraposite:

\begin{obs} \label{obs:red-solid-red}
	If $(C, \red[C])$ simulates $(B, \red[B])$,
	$(B, \red[B])$ simulates $(A, \red[A])$ conservatively,
	and the induced simulation of $(A, \red[A])$ by $(C, \red[C])$
		is non-conservative,
	then the simulation of $(B, \red[B])$ by $(C, \red[C])$
		is non-conservative.
	\emph{In the diagram: if $\Lambda_2$
	is below \colorbox{cno}{$\Lambda_1$} and they are related with a 
	{\color{cno}\underline{\normalcolor solid link}},
	then \colorbox{cno}{$\Lambda_2$}.}
\end{obs}

\paragraph{(Non-)conservativity results for uniform linear approximations}

In \cref{sec:uniformity}, we took advantage of
the notion of \emph{uniformity} introduced in Ehrhard and Regnier's
seminal work \cite{EhrhardRegnier08}
as a characterisation of actual approximants of λ-terms
out of all resource terms,
by porting it from resource terms to the resource reduction:
the uniform resource reduction $\cohrred$ restricts $\Lrred$ 
by acting on uniform sums of resource terms
and by reducing them in a uniform way.
Thanks to this construction, completed with the introduction
of a 001-infinitary closure $\cohrredi$ of $\cohrred$,
we were able to obtain a conservative simulation
of the 001-infinitary λ-calculus \emph{via} the Taylor expansion
(\cref{thm:conserv-cohrredi-bredi}).

Again, the two diagrams below describe which of the various
λ($\bot$)-calculi are simulated \colorbox{cyes}{conservatively},
\colorbox{cno}{non-conservatively},
or \colorbox{cout}{not simulated\vphantom{y}} at all,
respectively by $(\irsums,\cohrreds)$ and by $(\irsums,\cohrredi)$.

\begin{figure}[H]
\begin{widecenter}\footnotesize
\begin{tikzpicture}[x=1.5cm,y=1.2cm]
\node[node,nyes] 	(l-bs) 		at (2,4) {$(\lc, 	\breds)$};
\node[node,nyes]	(lb-bs) 	at (1,3) {$(\lbc, 	\breds)$};
\node[node,nyes]	(li-bs) 	at (3,3) {$(\lci, 	\breds)$};
\node[node,nout]	(lb-bbs) 	at (0,2) {$(\lbc, 	\bbotreds)$};
\node[nsim,nyes]	(lbi-bs) 	at (2,2) {$(\lbci, 	\breds)$};
\node[node,nout] 	(li-bi) 	at (4,2) {$(\lci, 	\bredi)$};
\node[node,nout]	(lbi-bbs) 	at (1,1) {$(\lbci, 	\bbotreds)$};
\node[node,nout] 	(lbi-bi) 	at (3,1) {$(\lbci, 	\bredi)$};
\node[node,nout] 	(lbi-bbi) 	at (2,0) {$(\lbci, 	\bbotredi)$};
\draw[l,lyes]	(l-bs) 		-- 				(lb-bs);
\draw[l,lyes]	(l-bs) 		-- 				(li-bs);
\draw[l,lyes]	(lb-bs) 	-- 				(lbi-bs);
\draw[l,lyes] 	(li-bs) 	-- 				(lbi-bs);
\draw[lnc,lout]	(lb-bs) 	-- 				(lb-bbs);
\draw[l,lout] 	(lb-bbs) 	-- 				(lbi-bbs);
\draw[lnc,lout]	(lbi-bs) 	-- 				(lbi-bbs);
\draw[lnc,lout]	(li-bs) 	-- 				(li-bi);
\draw[lnc,lout]	(lbi-bs) 	-- 				(lbi-bi);
\draw[l,lout]	(li-bi) 	-- 				(lbi-bi);
\draw[lnc,lout]	(lbi-bbs) 	-- 				(lbi-bbi);
\draw[lnc,lout]	(lbi-bi) 	-- 				(lbi-bbi);
\draw[l,lout]	(l-bs) 		to [bend right]	(lb-bbs);
\draw[l,lout]	(l-bs) 		to [bend left]	(li-bi);
\draw[l,lout]	(lb-bbs)	to [bend left]	(lbi-bbi);
\draw[l,lout]	(li-bi)		to [bend right]	(lbi-bbi);
\draw[l,lout]	(lb-bs)		to [bend left]	(lbi-bi);
\draw[l,lout]	(li-bs)		to [bend right]	(lbi-bbs);
\node[below=0 of lbi-bs, tyes] 	{\cref{cor:conserv-cohrred-bred}};
\node[above=0 of lbi-bs, tsim] 
	{\cref{thm:cohrred-simul-bred} \cite{CerdaPhD}};
\end{tikzpicture}
\qquad
\begin{tikzpicture}[x=1.5cm,y=1.2cm]
\node[node,nyes] 	(l-bs) 		at (2,4) {$(\lc, 	\breds)$};
\node[node,nyes]	(lb-bs) 	at (1,3) {$(\lbc, 	\breds)$};
\node[node,nno]		(li-bs) 	at (3,3) {$(\lci, 	\breds)$};
\node[node,nout]	(lb-bbs) 	at (0,2) {$(\lbc, 	\bbotreds)$};
\node[node,nno]		(lbi-bs) 	at (2,2) {$(\lbci, 	\breds)$};
\node[node,nyes] 	(li-bi) 	at (4,2) {$(\lci, 	\bredi)$};
\node[node,nout]	(lbi-bbs) 	at (1,1) {$(\lbci, 	\bbotreds)$};
\node[nsim,nyes] 	(lbi-bi) 	at (3,1) {$(\lbci, 	\bredi)$};
\node[node,nout] 	(lbi-bbi) 	at (2,0) {$(\lbci, 	\bbotredi)$};
\draw[l,lyes]	(l-bs) 		-- 				(lb-bs);
\draw[l]		(l-bs) 		-- 				(li-bs);
\draw[l]		(lb-bs) 	-- 				(lbi-bs);
\draw[l,lno] 	(li-bs) 	-- 				(lbi-bs);
\draw[lnc,lout]	(lb-bs) 	-- 				(lb-bbs);
\draw[l,lout] 	(lb-bbs) 	-- 				(lbi-bbs);
\draw[lnc,lout]	(lbi-bs) 	-- 				(lbi-bbs);
\draw[lnc,lno]	(li-bs) 	-- 				(li-bi);
\draw[lnc,lno]	(lbi-bs) 	-- 				(lbi-bi);
\draw[l,lyes]	(li-bi) 	-- 				(lbi-bi);
\draw[lnc,lout]	(lbi-bbs) 	-- 				(lbi-bbi);
\draw[lnc,lout]	(lbi-bi) 	-- 				(lbi-bbi);
\draw[l,lout]	(l-bs) 		to [bend right]	(lb-bbs);
\draw[l,lyes]	(l-bs) 		to [bend left]	(li-bi);
\draw[l,lout]	(lb-bbs)	to [bend left]	(lbi-bbi);
\draw[l,lout]	(li-bi)		to [bend right]	(lbi-bbi);
\draw[l,lyes]	(lb-bs)		to [bend left]	(lbi-bi);
\draw[l,lout]	(li-bs)		to [bend right]	(lbi-bbs);
\node[below=0 of lbi-bi, tyes] 	{\cref{thm:conserv-cohrredi-bredi}};
\node[above=0 of lbi-bi, tsim] 	{\cref{cor:cohrredi-simul-bredi}};
\end{tikzpicture}%
\end{widecenter}
\end{figure}

As in the previous paragraph, some results have been deduced from the others using
\cref{obs:green-solid-green,obs:red-solid-red},
or the following additional contraposite:

\begin{obs} \label{obs:green-dashed-red}
	If $(C, \red[C])$ simulates $(B, \red[B])$ conservatively
	and $(B, \red[B])$ simulates $(A, \red[A])$ non-conservatively,
	then $(C, \red[C])$ simulates $(A, \red[A])$ non-conservatively.
	\emph{In the diagram: if \colorbox{cyes}{$\Lambda_2$}
	is below $\Lambda_1$ and they are related with a 
	{\color{cno}\tunderdash{\normalcolor dashed link}},
	then \colorbox{cno}{$\Lambda_1$}.}
\end{obs}

In this restriction of $\Lrred$ to uniform resource reduction,
one thing was lost: the ability to simulate not only β-reductions,
but also β$\bot$-reductions, which is a key feature
of the linear approximation.
Therefore we slightly adapted our definitions to re-introduce
the behaviour of $\Lrred$ in the uniform resource reductions,
but limited to what is needed to simulate $\bot$-reduction steps.
The two diagrams below describe which of the various λ($\bot$)-calculi
are simulated by these adapted reduction systems,
namely $(\irsums,\cohrbotreds)$ and $(\irsums,\cohrbotredi)$,
respectively.

\begin{figure}[H]
\begin{widecenter}\footnotesize
\begin{tikzpicture}[x=1.5cm,y=1.2cm,baseline]
\node[node,nyes] 	(l-bs) 		at (2,4) {$(\lc, 	\breds)$};
\node[node,nno]		(lb-bs) 	at (1,3) {$(\lbc, 	\breds)$};
\node[node,nyes]	(li-bs) 	at (3,3) {$(\lci, 	\breds)$};
\node[node,nyes]	(lb-bbs) 	at (0,2) {$(\lbc, 	\bbotreds)$};
\node[node,nno]		(lbi-bs) 	at (2,2) {$(\lbci, 	\breds)$};
\node[node,nout] 	(li-bi) 	at (4,2) {$(\lci, 	\bredi)$};
\node[nsim,nyes]	(lbi-bbs) 	at (1,1) {$(\lbci, 	\bbotreds)$};
\node[node,nout] 	(lbi-bi) 	at (3,1) {$(\lbci, 	\bredi)$};
\node[node,nout,anchor=base] (lbi-bbi) at (2,0) {$(\lbci, \bbotredi)$};
\draw[l]		(l-bs) 		-- 				(lb-bs);
\draw[l,lyes]	(l-bs) 		-- 				(li-bs);
\draw[l,lno]	(lb-bs) 	-- 				(lbi-bs);
\draw[l] 		(li-bs) 	-- 				(lbi-bs);
\draw[lnc,lno]	(lb-bs) 	-- 				(lb-bbs);
\draw[l,lyes] 	(lb-bbs) 	-- 				(lbi-bbs);
\draw[lnc,lno]	(lbi-bs) 	-- 				(lbi-bbs);
\draw[lnc,lout]	(li-bs) 	-- 				(li-bi);
\draw[lnc,lout]	(lbi-bs) 	-- 				(lbi-bi);
\draw[l,lout]	(li-bi) 	-- 				(lbi-bi);
\draw[lnc,lout]	(lbi-bbs) 	-- 				(lbi-bbi);
\draw[lnc,lout]	(lbi-bi) 	-- 				(lbi-bbi);
\draw[l,lyes]	(l-bs) 		to [bend right]	(lb-bbs);
\draw[l,lout]	(l-bs) 		to [bend left]	(li-bi);
\draw[l,lout]	(lb-bbs)	to [bend left]	(lbi-bbi);
\draw[l,lout]	(li-bi)		to [bend right]	(lbi-bbi);
\draw[l,lout]	(lb-bs)		to [bend left]	(lbi-bi);
\draw[l,lyes]	(li-bs)		to [bend right]	(lbi-bbs);
\node[below=0 of lbi-bbs, tyes] 	{\cref{thm:conserv-cohrbotred-botred}};
\node[above=0 of lbi-bbs, tsim] 	{\cref{thm:cohrbotred-simul-bbotred}};
\end{tikzpicture}
\qquad
\begin{tikzpicture}[x=1.5cm,y=1.2cm,baseline]
\node[node,nyes] 	(l-bs) 		at (2,4) {$(\lc, 	\breds)$};
\node[node,nno]		(lb-bs) 	at (1,3) {$(\lbc, 	\breds)$};
\node[node,nno]		(li-bs) 	at (3,3) {$(\lci, 	\breds)$};
\node[node,nyes]	(lb-bbs) 	at (0,2) {$(\lbc, 	\bbotreds)$};
\node[node,nno]		(lbi-bs) 	at (2,2) {$(\lbci, 	\breds)$};
\node[node,nyes] 	(li-bi) 	at (4,2) {$(\lci, 	\bredi)$};
\node[node,nno]		(lbi-bbs) 	at (1,1) {$(\lbci, 	\bbotreds)$};
\node[node,nno] 	(lbi-bi) 	at (3,1) {$(\lbci, 	\bredi)$};
\node[nsim,nyes,anchor=base] (lbi-bbi) at (2,0) {$(\lbci, \bbotredi)$};
\draw[l]		(l-bs) 		-- 				(lb-bs);
\draw[l]		(l-bs) 		-- 				(li-bs);
\draw[l,lno]	(lb-bs) 	-- 				(lbi-bs);
\draw[l,lno] 	(li-bs) 	-- 				(lbi-bs);
\draw[lnc,lno]	(lb-bs) 	-- 				(lb-bbs);
\draw[l]	 	(lb-bbs) 	-- 				(lbi-bbs);
\draw[lnc]		(lbi-bs) 	-- 				(lbi-bbs);
\draw[lnc,lno]	(li-bs) 	-- 				(li-bi);
\draw[lnc]		(lbi-bs) 	-- 				(lbi-bi);
\draw[l]		(li-bi) 	-- 				(lbi-bi);
\draw[lnc,lno]	(lbi-bbs) 	-- 				(lbi-bbi);
\draw[lnc,lno]	(lbi-bi) 	-- 				(lbi-bbi);
\draw[l,lyes]	(l-bs) 		to [bend right]	(lb-bbs);
\draw[l,lyes]	(l-bs) 		to [bend left]	(li-bi);
\draw[l,lyes]	(lb-bbs)	to [bend left]	(lbi-bbi);
\draw[l,lyes]	(li-bi)		to [bend right]	(lbi-bbi);
\draw[l]		(lb-bs)		to [bend left]	(lbi-bi);
\draw[l]		(li-bs)		to [bend right]	(lbi-bbs);
\node[below=0 of lbi-bbi, tyes] 	{\cref{thm:conserv-cohrbotredi-botredi}};
\node[above=0 of lbi-bbi, tsim] 	{\cref{cor:cohrbotredi-simul-bbotredi}};
\end{tikzpicture}%
\end{widecenter}
\end{figure}

\paragraph{Further perspectives}

The picture is almost complete:
the uniform resource reductions, which we introduced
for their very constrained behaviour to ensure conservativity properties,
have been exhaustively studied in this paper;
as for the original reduction $\Lrred$ on sums of resource terms,
it is much more uncontrollable
(as the complexity of the Accordion demonstrates)
but for almost all λ($\bot$)-calculi we managed to discriminate
whether they were simulated conservatively by $\Lrred$ or not.
However, in the figure of \cpageref{fig:simulations-by-Lrred}
one \colorbox{cbase}{indeterminate} simulation remains...
Indeed, in the case of $(\lbc,\bbotreds)$,
no simple argument allows to conclude that the simulation
is conservative or not.
The \enquote{mashup} technique from \cref{sec:mashup}
is also inapplicable to this case,
because \cref{lem:mashup-5tay} fails in the presence of~$\bot$.
However we believe that this is a conservative case,
although we leave it for future work.

\begin{conj} \label{conj:conserv-Lrred-lbc-bbotreds}
	For all $M,N \in \lbc$, if $\Tay(M) \Lrred \Tay(N)$
	then $M \bbotreds N$.
\end{conj}

Beyond that,
the question naturally arises whether this approach is transferrable
to the richer λ-calculi already endowed with a linear approximation
(as listed in the introduction).
This remains unclear, since most of these settings are
non-uniform, \ie it is not true any more that $\Tay(M) \coh \Tay(M)$
in general. Investigating how existing techniques used to tame
non-uniformity, \emph{e.g.} in \cite{Vaux19}, can be exploited
to address the conservativity problem in richer settings,
remains an open line of research.

%% --------------------------------------------------------------------------

%\section*{Acknowledgments}

\printbibliography

\end{document}